\documentclass[11pt]{article}
\usepackage{fullpage}
\usepackage{amsmath}
\usepackage{amssymb}
\usepackage[dvips]{epsfig}
\usepackage{color}

\usepackage{type1cm}        % activate if the above 3 fonts are
\usepackage{makeidx}         % allows index generation
\usepackage{graphicx}        % standard LaTeX graphics tool
 \usepackage{stackrel}

\usepackage{newtxtext}       % 
\usepackage[varvw]{newtxmath}       % selects Times Roman as basic font

\usepackage{wrapfig}
\usepackage{fullpage}
\usepackage{dcolumn}

\usepackage{bm}
\usepackage{color}
\usepackage{url}

\usepackage{placeins}
\usepackage{inputenc}
\usepackage{ulem}
\usepackage{upgreek}
\def\##1{{\bf #1}}
\def\=#1{\underline{\underline #1}}
\def\^#1{{\rR}eve{#1}}

\def\.{\mbox{ \tiny{$^\bullet$} }}

\def\ux{\hat{\#x}}
\def\uy{\hat{\#y}}
\def\uz{\hat{\#z}}
 
\def\ur{\hat{\#r}}
\def\uphi{\hat{\mbox{\boldmath$\phi$}}}
\def\utheta{\hat{\mbox{\boldmath$\theta$}}}

\def\muo{\mu_{\scriptscriptstyle 0}}
\def\epso{\eps_{\scriptscriptstyle 0}}
\def\etao{\eta_{\scriptscriptstyle 0}}
\def\lambdao{\lambda_{\scriptscriptstyle 0}}
\def\ko{k_{\scriptscriptstyle 0}}

\def\eps{\varepsilon}
\def\epsa{\eps_{\rm a}}
\def\epsb{\eps_{\rm b}}
\def\epsc{\eps_{\rm c}}

\def\le{\left(}
\def\ri{\right)}
\def\les{\left[}
\def\ris{\right]}
\def\lec{\left\{}
\def\ric{\right\}}

\def\deg{^\circ}

\def\aal{a_{\rm L}}
\def\aar{a_{\rm R}}
\def\bbl{r_{\rm L}}
\def\bbr{r_{\rm R}}
\def\ccl{t_{\rm L}}
\def\ccr{t_{\rm R}}

\def\aas{a_{\rm s}}
\def\aap{a_{\rm p}}
\def\bbs{r_{\rm s}}
\def\bbp{r_{\rm p}}
\def\ccs{t_{\rm s}}
\def\ccp{t_{\rm p}}

\def\thetainc{\theta_{\rm inc}}
\def\phiinc{\phi_{\rm inc}}

\def\szero{s_0}
\def\sone{s_1}
\def\stwo{s_2}
\def\sthree{s_3}

\def\sinczero{s_0^{\rm inc}}
\def\sincone{s_1^{\rm inc}}
\def\sinctwo{s_2^{\rm inc}}
\def\sincthree{s_3^{\rm inc}}

\def\alphainc{\alpha_{\rm inc}}
\def\betainc{\beta_{\rm inc}}

\def\strazero{s_0^{\rm tr}}
\def\straone{s_1^{\rm tr}}
\def\stratwo{s_2^{\rm tr}}
\def\strathree{s_3^{\rm tr}}

\def\Arg{{\rm Arg}}

\def\sp{\#s}
\def\spinc{\#s_{\rm inc}}
\def\pinc{\#p_{\rm inc}}
\def\pref{\#p_{\rm ref}}

\def\winc{w_{\rm inc}}

\def\Ei{\#E_{\rm inc}(\#r)}

\def\Er{\#E_{\rm ref}(\#r)}

\def\Et{\#E_{\rm tr}(\#r)}

\def\Re{{\rm Re}}
\def\Im{{\rm Im}}

\def\psis{\underline{\psi}^{\rm s}}

\def\bpsis{\les\underline{\psi}^{\rm s}\ris}
\def\bpsia{\les\underline{\psi}^{\rm a}\ris}
\def\bpsism{\les\underline{\psi}^{\rm s}_m\ris}
\def\bpsisn{\les\underline{\psi}^{\rm s}_n\ris}
\def\bpsiam{\les\underline{\psi}^{\rm a}_m\ris}
\def\bpsian{\les\underline{\psi}^{\rm a}_n\ris}

\def\Psis{\Psi^{\rm s}}
\def\Psia{\Psi^{\rm a}}

\def\tpsi{{\tilde \psi}}
\def\bpsipq{\les\underline{\tpsi}^{(p,q)}\ris}
\def\bpsipqm{\les\underline{\tpsi}^{(p,q)}_m\ris}
\def\bpsipqn{\les\underline{\tpsi}^{(p,q)}_n\ris}
\def\bpsizeroq{\les\underline{\tpsi}^{(0,q)}\ris}

\def\bpsipqinc{\les\underline{\tpsi}^{(p,q)}_{\rm inc}\ris}
\def\bpsipqref{\les\underline{\tpsi}^{(p,q)}_{\rm ref}\ris}
\def\bpsipqtra{\les\underline{\tpsi}^{(p,q)}_{\rm tr}\ris}
\def\bpsipqsca{\les\underline{\tpsi}^{(p,q)}_{\rm sca}(\ur)\ris}

\def\tupsilon{{\tilde\upsilon}}
\def\tPsi{{\tilde\Psi}}
\def\Psipqmn{\tPsi^{\left(p,q\right)}_{mn}}
\def\Psipqmm{\tPsi^{\left(p,q\right)}_{mm}}
\def\Psipqnm{\tPsi^{\left(p,q\right)}_{nm}}

\def\Psipqref{\tPsi^{\left(p,q\right)}_{\rm ref}}
\def\Psipqtra{\tPsi^{\left(p,q\right)}_{\rm tr}}
\def\Psipqsca{\tPsi^{\left(p,q\right)}_{\rm sca}(\ur)}

\def\Vint{{\cal V}_{\rm int}}

\def\Vext{{\cal V}_{\rm ext}}

\def\Dmn{D_{\rm mn}}
\def\eomn{_{\stackrel[{\rm o}]{\rm e}{}{_{\rm mn}}}}
\def\smn{_{\rm smn}}

\def\sca{_{\rm sca}}
 
\def\Thetasca{\Theta_{\rm sca}}
\def\Phisca{\Phi_{\rm sca}}

\def\pinm{\pi_{\rm n}^{\rm m}}
\def\taunm{\tau_{\rm n}^{\rm m}}
\def\one{^{(1)}}
\def\three{^{(3)}}

\def\epsr{\eps_{\rm r}}
\def\mur{\mu_{\rm r}}
\def\QD{Q_{\rm D}}

\begin{document}
 
 \begin{center}

\LARGE{ {\bf The double-indexed geometric phase for electromagnetics 
}}
\end{center}
\begin{center}
\vspace{10mm} \large
%\today

 {Akhlesh  Lakhtakia}\footnote{E--mail: akhlesh@psu.edu}\\
  \textit{ NanoMM~---~Nanoengineered Metamaterials Group\\ Department of Engineering Science and Mechanics\\
The Pennsylvania State University, University Park, PA 16802--6812, USA}\\
and\\
 \textit{ School of Mathematics,
University of Edinburgh, Edinburgh EH9 3FD, UK}
 \vspace{3mm}\\
 {Tom G. Mackay}\footnote{E--mail: T.Mackay@ed.ac.uk.}\\
 \textit{ School of Mathematics and
   Maxwell Institute for Mathematical Sciences\\
University of Edinburgh, Edinburgh EH9 3FD, UK}\\
and\\
  \textit{ NanoMM~---~Nanoengineered Metamaterials Group\\ Department of Engineering Science and Mechanics\\
The Pennsylvania State University, University Park, PA 16802--6812,
USA}

\normalsize

\end{center}

\begin{center}
\vspace{5mm} {\bf Abstract}
\end{center}

The double-indexed geometric phase (DIGP) based on a family of Poincar\'e spinors was formulated to extend the Pancharatnam phase, commonly used to describe
the evolution of the polarization state in an optical system relative to a reference polarization state. Analytical results establish the symmetries of the DIGP under reversal of 
the index $p\in\mathbb{R}$ for the latitude on the Poincar\'e sphere, the phase change induced by reversal of the index $q\in\mathbb{R}$
for the longitude on the same sphere, and the periodic dependence on $q$. For closed  loops on the Poincar\'e sphere, the DIGP can be interpreted geometrically through 
a $p$-dependent mapping of both geographic coordinates and the associated  solid angle subtended at the center of the sphere. 
Although the geometric phase is usually applied to
compare two plane waves propagating in the same direction in free space, the DIGP  is expected to contain information when comparing two non-co-propagating
plane waves, as demonstrated  by specular reflection by a planar thin film and far-zone scattering by a three-dimensional object. DIGP maps for specular
reflection and transmission by thin films resolve Bragg phenomenons, Fabry--P\'erot resonances, structural chirality, anisotropy, and defects 
with detail not evident in reflectance and transmittance maps. A map of  direction-dependent DIGP for plane-wave scattering by a three-dimensional object may
reveal strong polar and azimuthal structure absent from the map of differential scattering efficiency. Principal component analysis indicates that a collection of DIGP maps,
all calculated
from the same set of polarimetric measurements,
form a coordinated data family rather than a collection of independent observables.
The DIGP concept   offers a 
complementary  framework for polarization-state-resolved characterization and inverse scattering.

%\maketitle

\section{Introduction}\label{sec:intro} 

In the broadest sense, a spinor is a complex vector whose transformation law under rotations is not the same as that of an ordinary vector or tensor: a $2\pi$ rotation  changes the sign of the spinor, and only a $4\pi$ rotation necessarily returns it to itself \cite{Cartan}. This double-valued behavior is a hallmark of the fundamental representation of SU(2), the group of all 2$\times$2 unitary matrixes with unit determinant.

In physics, spinors appear in multiple but closely related contexts, the most familiar being relativistic Dirac spinors, non-relativistic Pauli spinors, and  Poincar\'e or polarization spinors in optics.  Although the physical interpretations differ, the common mathematical core is the same: a complex two-component   or four-component
state object, projective phase equivalence, and symmetry under rotations represented by unitary transformations. This common structure is what makes spinors so powerful across physics, and it is precisely this structure that underlies the concept of geometric phase \cite{Shapere}.

The Dirac spinor is the most general and physically richest of the canonical spinors in relativistic quantum theory \cite{Rosen}. It is a four-component complex field that transforms under the Lorentz group and satisfies the Dirac equation, thereby providing the quantum description of spin-$\tfrac12$ particles such as electrons and positrons. Its chief properties include relativistic covariance, the coexistence of particle and antiparticle degrees of freedom, and the ability to encode spin and chirality within a single field-theoretic object.

The Pauli spinor is the two-component counterpart appropriate to nonrelativistic spin-$\tfrac12$ systems \cite{Lounesto}. In this setting, a spin state is represented by a complex two-component vector whose evolution is governed by the Pauli equation, with the Pauli matrixes acting as the generators of spin rotations. The Pauli spinor is central to the description of magnetic resonance, spin dynamics in atomic and condensed-matter systems, and many aspects of quantum information. Its chief mathematical features are normalization, phase redundancy, SU(2) rotational covariance, and the fact that pure spin states correspond to points on the Bloch sphere. 

In optics, which concerns us most in this paper, the analogous object is the Poincar\'e spinor. Its two components represent orthogonal polarization amplitudes, often taken in a circular or linear basis, and the spinor encodes the complete state of a monochromatic fully polarized plane wave propagating in free space. Its chief properties mirror those of other two-state spinors: it is defined up to an overall phase, it transforms under SU(2)-type rotations, and its physical content is naturally visualized on the Poincar\'e sphere
\cite{BW1998}. The geometry of polarization state then becomes a geometry of state space, with relative phase, ellipticity, and handedness, all represented in a unified way. This formalism  provides the natural language for the geometric phase of polarized light \cite{Pancha1,Gori1997,Bhandari,GV}.

The Poincar\'e spinor is compared with the Pauli and Dirac spinors in Table~\ref{table1} for the interested reader, but we now press onwards here  for
electromagnetics.

%%%%%%%%%%%%%%%%%%%% Table 1 begins %%%%%%%%%%%%%%%%%%%%
\newpage
\begin{table}[ht]
\caption{Comparison of Poincar\'e, Pauli, and Dirac spinors in terms of mathematical structure and physical interpretation.}
\label{table1}
\centering
\renewcommand{\arraystretch}{1.0}
\begin{tabular}{|p{3.4cm}|p{3.8cm}|p{3.8cm}|p{3.8cm}|}
\hline
\textbf{Property} & \textbf{Poincar\'e spinor} & \textbf{Pauli spinor} & \textbf{Dirac spinor} \\
\hline
\textbf{Typical object} & Polarization state of light & Non-relativistic spin-$\tfrac12$ state & Relativistic spin-$\tfrac12$ field \\
\hline
\textbf{Vector space} & $\mathbb C^2$ & $\mathbb C^2$ & $\mathbb C^4$ \\
\hline
\textbf{Standard form} $[{\underline \psi}]$& $\begin{bmatrix}\psi_a\\ \psi_b\end{bmatrix}$ & $\begin{bmatrix}\psi_a\\ \psi_b\end{bmatrix}$ & $\begin{bmatrix}\psi_a\\ \psi_b\\ \psi_c\\ \psi_d\end{bmatrix}$ \\
\hline
\textbf{Normalization} & $|\psi_a|^2+|\psi_b|^2=1$ & $|\psi_a|^2+|\psi_b|^2=1$ & Via bilinears  \\
\hline
\textbf{Relative phase determines} & Ellipticity/handedness & Spin superposition & Chiral/particle--antiparticle content \\
\hline
\textbf{Geometric representation} & Poincar\'e sphere & Bloch sphere & No simple 2-sphere representation \\
\hline
\textbf{Basis choice} & Usually $L/R$ circular or $H/V$ linear polarization & Usually spin-up/spin-down along an axis & Dirac, Weyl, etc. representations \\
\hline
\textbf{Symmetry group} & $SU(2)$ on polarization states & $SU(2)$ spin rotations & proper orthochronous Lorentz group \\
\hline
\textbf{Rotation behavior} & $2\pi$ rotation gives sign flip   & $2\pi$ rotation gives sign flip & $2\pi$ rotation gives sign flip \\
\hline
\textbf{Basic parametrization} & $\begin{bmatrix}\cos(\vartheta/2)\\ \sin(\vartheta/2)e^{i\varphi}\end{bmatrix}$ up to phase & $\begin{bmatrix}\cos(\vartheta/2)\\ \sin(\vartheta/2)e^{i\varphi}\end{bmatrix}$ for two-level spin state & No simple spherical parametrization \\
\hline
\textbf{Key matrixes} & Jones matrixes, Pauli matrixes in polarization basis & Pauli matrixes $[\boldsymbol{\sigma}_{x,y,z}]$ & Gamma matrixes \\
\hline
\textbf{Algebraic structure} & Projective $SU(2)$ geometry & $SU(2)$ spin algebra & Clifford algebra  \\
\hline
\textbf{Transformations} & Jones/$SU(2)$ rotations of polarization state & Spin rotations & Lorentz-covariant spinor transformations \\
\hline
\textbf{Observable content} & Stokes parameters  & Spin expectation values  & Bilinears   \\
\hline
\textbf{Evolution equation} & Jones calculus & Pauli equation & Dirac equation \\
\hline
\textbf{Geometric phase} & Pancharatnam (or Pancharatnam--Berry) phase & Can appear in spin transport / adiabatic evolution & Appears in relativistic transport but not the usual central focus \\
\hline
\textbf{Physical domain} & Polarization optics  & Non-relativisitic quantum mechanics & Relativistic quantum mech\-anics  \\
\hline
\textbf{Typical use} & Describing  polarization state & Describing spin precession and two-level states & Describing electrons, positrons, and relativistic fermions \\
\hline
\end{tabular}
\end{table}
%%%%%%%%%%%%%%%%%%%% Table 1 ends %%%%%%%%%%%%%%%%%%%%

Geometric phase entered electromagnetics   with Pancharatnam’s 1956 work on the interference of polarized beams, in which he showed that the relative phase between two non-identical polarization states is fixed by their overlap on the Poincar\'e sphere \cite{Pancha1}. This phase is not a conventional dynamical phase. Instead, it is geometric in nature and depends on the path taken by the polarization state through projective state space. Independently, Berry later proposed the same
concept for adiabatically evolving quantum systems and demonstrated that a cyclic evolution produces a phase equal to half the solid angle enclosed by the path in parameter space \cite{Berry1984}. In optics, this result clarified the role of geometric phase in polarization-state transport and established a deep connection between spinor geometry, holonomy, and the Poincar\'e sphere.

The optics research community rapidly recognized that polarization states provide a natural physical realization of projective two-state geometry \cite{Bhandari}. The   Poincar\'e sphere made the geometry visually transparent: linear, circular, and elliptical polarizations correspond to points on that sphere, and closed loops on that sphere generate a geometric phase proportional to the enclosed solid angle.  

A major advance came from experiments on polarized light in curved optical fibers. Tomita and Chiao demonstrated in 1986 that an optical beam transported through a helically wound fiber acquires a measurable geometric phase, confirming the Pancharatnam  phase \cite{footnote1}
%\footnote{The Pancharatnam phase is also called
%the Pancharatnam--Berry phase in optics literature.} 
for  the polarization state  \cite{Tomita}. Their result was an important milestone because it showed that geometric phase is not merely a formal construct but an observable optical effect. Experimental confirmations of the existence of the Pancharatnam phase resulting from the traversal
of a laser beam along a closed loop on the Poincar\'e
sphere came in 1988 \cite{BS1988,Chyba}. And research on geometric phase in optics took off \cite{Bhandari1994,Rao2026}.

The initial and final states are the same on a closed loop. Pancharatnam's conception had emerged simply from a comparison of two beams, not from the
initial and final states of the same beam. In 2010, 
van Dijk \textit{et al.}   reported the first measurement of the geometric phase arising from a non-cyclic transformation of the polarization state \cite{vanDijk}. Non-cyclic transformations are especially relevant to planar electromagnetic devices.
Recently, geometric-phase metasurfaces have demonstrated that rotating anisotropic meta-atoms can imprint helicity-dependent phase shifts, thereby enabling beam steering, holography, and diffusion-like scattering  \cite{Chen2016}.
Applications now span wavefront shaping, spin-controlled beam steering, holography, orbital-angular-momentum generation, polarization  splitting, and flat lenses, among others
\cite{ChenReview,YuReview,JishaReview,FarazReview}.

Can Pancharatnam's conception of the geometric phase imprinted on the Poincar\'e sphere be extended? Truly, his conception does not need the Poincar\'e sphere, his eponymous phase arising simply from an inner product of two column vectors belonging to $\mathbb{C}^2$. A similar idea underlies the electro-magnetic geometric phase
\cite{Vernon} and the Riemann--Silberstein phase \cite{Cheng} for a
spin-$1$ system (i.e., electromagnetic fields), both geometric phases
arising from an inner product of two column vectors belonging to $\mathbb{C}^4$. 

We present here the  double-indexed geometric phase (DIGP), of which the Pancharatnam phase is a special case. Conceptualization of the Pancharatnam
phase does not require
the Poincar\'e sphere, but that sphere is necessary to define  the DIGP. The DIGP is motivated by the possibility of extracting information about the scattering
object from it \cite{LakhtakiaAthens}, but the realization of that possibility is a future research direction. 

The plan of this paper is as follows: Section.~\ref{sec:PP} begins with the Poincar\'e sphere, establishes the symmetric Poincar\'e spinor,  retraces the steps taken by Pancharatnam to cast the interference of two co-propagating waves in terms of a geodesic angle and the geometric (Pancharatnam) phase, and relates the closed-loop geometric phase to the  associated  solid angle subtended at the center of the sphere. Section~\ref{sec:sg} introduces the family of double-indexed Poincaré spinors and establishes the  symmetric and asymmetric Poincaré spinors  as special cases. Section~\ref{sec:DIGP} defines the DIGP through the  double-indexed overlap and goes on to
establish its principal symmetries  under reversal of 
the index $p\in\mathbb{R}$ for the latitude on the Poincar\'e sphere, the phase change induced by reversal of the index $q\in\mathbb{R}$
for the longitude on the same sphere, and the periodic dependence on $q$. A geometric interpretation of the DIGP for closed loops
on the Poincar\'e sphere is also developed in terms of a solid angle. Section~\ref{sec:examples} applies the formalism to plane-wave scattering by planar thin films
and three-dimensional objects, in order to demonstrate the sensitivity of DIGP to  resonances, anisotropy, structural chirality, non-homogeneity, and
defects. The paper concludes with Sec.~\ref{sec:conc} proffering the DIGP as a complementary  framework for electromagnetic characterization and inverse scattering.

An $ \exp(-i \omega t)$  dependence on time $t$ is implicit, where $ \omega$ as the angular frequency and $i=\sqrt{-1}$.
With $ \epso$ and $\muo$, respectively,
denoting the permittivity and permeability  of free space,
the free-space wavenumber is denoted by $\ko 
= \omega \sqrt{\epso \muo}$,  and $\lambdao=2\pi/\ko$ is the free-space
wavelength. 
Vectors are in boldface
 and unit vectors are additionally decorated by a caret on top. Dyadics \cite{Chen} are double underlined.
 Column vectors are underlined and enclosed in square brackets. Matrixes are in boldface and enclosed in square brackets. The asterisk $(^\ast)$ denotes the complex conjugate
 and the dagger $(^\dag)$ denotes the conjugate transpose. The Cartesian coordinate system $(x,y,z)$ with unit vectors $\ux$, $\uy$, and $\uz$ is adopted
 along with the spherical coordinate system $(r,\theta,\phi)$ with unit vectors $\ur$, $\utheta$, and $\uphi$.

\section{The Poincar\'e sphere and the Pancharatnam phase}\label{sec:PP}

\subsection{Plane-wave representation on the Poincar\'e sphere}
The electric  field phasor associated with a uniform plane
wave propagating in free space (i.e., vacuum) at an angle $\theta \in[0,\pi/2)$  
with respect to the  ${\pm}z$ axis and at an angle $\phi\in[0,2\pi)$  with respect
to the $x$ axis in the $xy$ plane, is   represented as \cite{STFbook}
\begin{subequations}
\begin{eqnarray}
\label{eqE-lin}
\#E^\pm(\#r)&=& \le \aas^\pm \sp + \aap^\pm\#p_\pm\ri  \exp{\les i\kappa \le x\cos\phi+ y\sin\phi \ri
\ris} \exp{\left({\pm}i\ko z{\cos\theta}\right)}\\
\nonumber
&=&
\pm\les \frac{\le i\sp - \#p_\pm \ri}{\sqrt{2}} \, \aal^\pm -\, \frac{\le i\sp +\#p_\pm
\ri}{\sqrt{2}} \, \aar^\pm \ris
\exp{\les i\kappa \le x\cos\phi + y\sin\phi \ri
\ris}\\
 && \times \exp{\left({\pm}i\ko z{\cos\theta}\right)}\,,
 \label{eqE-circ}
\end{eqnarray}
\end{subequations}
with
 \begin{equation}
 \left.\begin{array}{l}
\kappa =
\ko\sin{\theta}\\[5pt]
\sp=-\ux\sin\phi + \uy \cos\phi \\[5pt]
\#p_\pm=\mp\le \ux \cos\phi + \uy \sin\phi \ri \cos{\theta} 
 + \uz \sin{\theta}
\end{array}
\right\}
\, .
\end{equation}
The amplitudes  of the perpendicular- and parallel-polarized
components, respectively, are denoted by $\aas^\pm\in\mathbb{C}$ and $\aap^\pm\in\mathbb{C}$ in Eq.~(\ref{eqE-lin}).
The amplitudes  of the
left-  and  the 
 right-circularly polarized (LCP and RCP)  components  are denoted by 
$\aal^\pm\in\mathbb{C}$ and
$\aar^\pm\in\mathbb{C}$, respectively, in Eq.~(\ref{eqE-circ}).  The upper signs are to be used for the up-going plane wave
and the lower signs for the down-going plane wave, with respect to the $+z$ axis.

The Stokes parameters of the  plane wave are given by \cite{Jackson}
\begin{equation}
\label{Stokes}
\left.\begin{array}{l}
\szero = \vert\aal^\pm\vert^2+\vert\aar^\pm\vert^2 =\vert\aas^\pm\vert^2+\vert\aap^\pm\vert^2
\\[5pt]
\sone =2\,\Re\left(\aal^\pm\,\aar^{\pm\ast}\right)=\vert\aap^\pm\vert^2-\vert\aas^\pm\vert^2
\\[5pt]
\stwo  =2\,\Im\left(\aal^\pm\,\aar^{\pm\ast}\right)=2\,\Re\left(\aas^\pm\,\aap^{\pm\ast}\right)
\\[5pt]
\sthree  = \vert\aar^\pm\vert^2-\vert\aal^\pm\vert^2=2\,\Im\left(\aas^\pm\,\aap^{\pm\ast}\right)
\end{array}
\right\}\,.
\end{equation} 
These parameters satisfy the condition
\begin{equation}
\label{eqPS}
(\sone/\szero)^2+(\stwo/\szero)^2+(\sthree/\szero)^2=1\,,
\end{equation}
which is the equation  of a unit-radius sphere called the Poincar\'e sphere after its originator \cite{Kahr}.
Accordingly, the plane wave can be located jointly by the longitude $\alpha\in[0,2\pi)$  and the latitude $\beta\in[-\pi/2,\pi/2]$  defined through the relations
\cite{Poincare,Deschamps}
\begin{equation}
\label{def-alphabeta}
\left.\begin{array}{l}
\sone=\szero\, \cos\beta\, \cos\alpha
\\[5pt]
\stwo=\szero\, \cos\beta\, \sin\alpha
\\[5pt]
\sthree=\szero\, \sin\beta 
\end{array}
\right\}\,.
\end{equation}

The  angles $\alpha$ and $\beta$     appear in the  symmetric Poincar\'e spinor
\begin{equation}
\label{def-PS-0}
\bpsis=\les
\begin{array}{c}
\cos\left(\frac{\pi}{4}-\frac{\beta}{2}\right)\exp\left(-i\alpha/2\right)
\\[5pt]
\sin\left(\frac{\pi}{4}-\frac{\beta}{2}\right)\exp\left(i\alpha/2\right)
\end{array}
\ris\,
\end{equation}
modeled
after the Jones vector \cite{Gori1997}, as well as in the asymmetric Poincar\'e spinor \cite{Das}
\begin{equation}
\label{def-PA-0}
\bpsia= \exp\left(i\alpha/2\right)\bpsis\,.
\end{equation}
Whereas the angle $\alpha$ appears in both elements of $\bpsis$, it appears in
only one element of $\bpsia$. Therefore, $\bpsis$ is \textsf{symmetric}
and  $\bpsia$   is \textsf{asymmetric} \cite{LakhtakiavdH}. 

The symmetric Poincar\'e spinor can be written as
\begin{equation}
\bpsis=\exp\lec-i\alpha[\boldsymbol{\sigma}_z]/2\ric\.\exp\lec-i \left(\frac{\pi}{2}-{\beta}\right)[\boldsymbol{\sigma}_y]/2\ric\.
\les\begin{array}{c}1\\0\end{array}\ris\,,
\end{equation}
where the Pauli matrixes
\begin{equation}
[\boldsymbol{\sigma}_y]=\les\begin{array}{ccc}0&&-i\\ i&&0\end{array}\ris
\qquad {\rm and}\qquad
[\boldsymbol{\sigma}_z]=\les\begin{array}{ccc}1&&0\\ 0&&-1\end{array}\ris\,.
\end{equation}
If the matrix $[\#U]$ is unitary with unit determinant, i.e.,
\begin{equation}
[\#U]= \les\begin{array}{ccc}u_1&&u_2\\ -u_2^\ast&&u_1^\ast\end{array}\ris\,,\qquad {\rm where}\qquad
\vert{u_1}\vert^2+\vert{u_2}\vert^2=1\,,
\end{equation}
then $[\#U]\.\bpsis$ is also a  spinor. Hence, $\bpsis$ is SU(2)-covariant. Furthermore,
in view of Eq.~\eqref{def-PA-0}, $\bpsia$ is also SU(2)-covariant.

\subsection{Pancharatnam's novelty}\label{novelty}
Pancharatnam   \cite{Pancha1}   studied the interference of two uniform plane waves, labelled $1$ and $2$,
propagating parallel to one another and endowed with the same frequency but possibly with different polarization states.  
The magnitude
of the time-averaged
Poynting vector of the two plane waves together turns out to be
\begin{equation}
P=P_1 + P_2 + 2\sqrt{P_1\,P_2}\, \cos\left(\frac{1}{2}\Delta_{12}\right) \cos\left(\Psis_{12}\right)\,,
\label{two}
\end{equation}
where $P_1$ is the analogous quantity for plane wave $1$ alone and $P_2$ for plane wave $2$ alone. 

Suppose
the respective labels are also attached to the locations of the two plane waves on the Poincar\'e sphere.
Then  $\Delta_{12}$
is defined as the angle subtended at the center of the Poincar\'e sphere by the geodesic (i.e., great circle) joining the locations
$1$ and $2$ on  that sphere. This angle is the \textsf{geodesic angle}.

The angle $\Psis_{12}$ was the novelty defined by Pancharatnam as \cite{Pancha1} ``\textit{the phase advance of one   polarised beam over another} (not necessarily in the same
state of polarisation) \textit{is the amount by which its phase must be retarded relative
to the second, in order that the intensity resulting from their mutual interference
may be a maximum}.'' This angle is the \textsf{Pancharatnam phase}.

If $M$ plane waves propagating parallel to one another and endowed with the same frequency but possibly with different polarization states were
to interfere, the analog of Eq.~(\ref{two}) was given by Pancharatnam \cite{Pancha1} as
\begin{equation}
P = \sum_{m=1}^{M}\,P_m + \sum_{m=1}^{M} \sum_{\stackrel{n=1}{n{\ne m}}}^{M}\sqrt{P_mP_n}\, \cos\left(\frac{1}{2}\Delta_{mn}\right) \cos\left(\Psis_{mn}\right)\,,
\end{equation}
confirming that his eponymous phase $\Psis_{mn}$ emerges solely from a comparison of plane waves labelled $m$ and $n$.

\subsection{Pancharatnam phase and geodesic angle}

With
\begin{equation}
\label{def-PS}
\bpsism=\les
\begin{array}{c}
\cos\left(\frac{\pi}{4}-\frac{\beta_m}{2}\right)\exp\left(-i\alpha_m/2\right)
\\[5pt]
\sin\left(\frac{\pi}{4}-\frac{\beta_m}{2}\right)\exp\left(i\alpha_m/2\right)
\end{array}
\ris\,
\end{equation}
following from Eq.~(\ref{def-PS-0})
and $\bpsisn$   defined similarly, the complex inner product
\begin{equation}
\upsilon_{mn}= {\bpsism}^\dag\.{\bpsisn}
\end{equation}
is called the Pancharatnam overlap \cite{footnote2}.
Its argument is the Pancharatnam phase, i.e.,
\begin{subequations}
\begin{eqnarray}
\label{GPS2-def}
\Psis_{mn}&=&{\rm Arg}\lec\upsilon_{mn}\ric
\\[5pt]
\nonumber
&=& {\rm Arg}\lec\cos\left(\frac{\alpha_m-\alpha_n}{2}\right)\cos\left(\frac{\beta_m-\beta_n}{2}\right)+\right.
\\[5pt]
&&\qquad \left.
i\sin\left(\frac{\alpha_m-\alpha_n}{2}\right)
\sin\left(\frac{\beta_m+\beta_n}{2}\right)\ric\,.
\label{GPS2}
\end{eqnarray}
\end{subequations}
All phase equalities involving $\Arg$ are understood modulo $(2\pi)$, unless a continuous branch is specified.
Also, $\Psis_{mn}$ is undefined when $\upsilon_{mn} = 0$, i.e., when the locations $m$ and $n$ are antipodal.

The unit vector joining the center of the Poincar\'e sphere to the location  of the $m^{\rm th}$ plane wave  on that
sphere is given by
\begin{equation}
\hat{\#r}_m= \left(\hat{\#u}_1\cos\alpha_m+\hat{\#u}_2\sin\alpha_m\right)\cos\beta_m +\hat{\#u}_3\sin\beta_m\,,
\end{equation}
where $\lec \hat{\#u}_1,\hat{\#u}_2,\hat{\#u}_3\ric$ is the triad of Cartesian unit vectors relevant to the Poincar\'e sphere
described by Eq.~\eqref{eqPS}.
Then the geodesic angle is
given by
\begin{eqnarray}
\label{def-GA}
\Delta_{mn}&=&\cos^{-1}\left(\hat{\#r}_m\.\hat{\#r}_n\right)
\\
&=&\cos^{-1}\left[\sin\beta_m\sin\beta_n+\cos\beta_m\cos\beta_n\cos\left(\alpha_m-\alpha_n\right)
\right]\,,
\end{eqnarray}
with $\hat{\#r}_n$ defined analogously to $\hat{\#r}_m$.

The geodesic angle and the Pancharatnam phase are related to each other by the identity
\begin{equation}
\upsilon_{mn}=\cos\left(\frac{1}{2}\Delta_{mn}\right) \exp\left(i\Psis_{mn}\right)\,.
\end{equation}

\subsection{Plane-wave evolution}
Suppose that a plane wave identified by location $1$ on the Poincar\'e sphere is incident on an optical system,
 a plane wave identified by location $2$ on the Poincar\'e sphere exits that system and is incident on a second
 optical system, and  a plane wave identified by location $3$ on the Poincar\'e sphere exits the second system.
Then, the intermediate
 plane wave may be said to have the  geometric phase
$\Psi^{\rm s}_{12}$ relative to the incident plane wave and  the exiting plane wave has the geometric
$\Psi^{\rm s}_{23}$ relative to the intermediate plane wave. Accordingly, the exiting plane wave
has the geometric phase
\begin{equation}
\label{def-123}
\Psi^{\rm s}_{123}= \Psi^{\rm s}_{12}+\Psi^{\rm s}_{23}  \pmod{2\pi}
\end{equation}
with the incident plane wave serving as the \textit{reference}.

Suppose instead that the plane wave located at $1$ is incident on a third optical system and the exiting plane wave
is located at $3$. Then, the exiting plane wave has the geometric phase $\Psi^{\rm s}_{13}$ relative the incident plane wave.
Pancharatnam showed that
\begin{equation}
\Psi^{\rm s}_{123}\ne\Psi^{\rm s}_{13}\,,
\end{equation}
indicating the path-dependence of the geometric phase.

The relationship between $\Psi^{\rm s}_{123}$ and $\Psi^{\rm s}_{13}$ was extracted from spherical geometry  by
Pancharatnam as follows. Consider the closed loop $1\to 2\to 3 \to 1$. The closed-loop overlap is the product of the three consecutive overlaps
is \cite{footnote3}
\begin{eqnarray}
\nonumber
\upsilon_{1231}&=&\upsilon_{12}\,\upsilon_{23}\,\upsilon_{31}
\\[5pt]
&=&\cos\left(\frac{1}{2}\Delta_{12}\right)\cos\left(\frac{1}{2}\Delta_{23}\right)\cos\left(\frac{1}{2}\Delta_{31}\right)
\exp\les{i\left(\Psi^{\rm s}_{12}+\Psi^{\rm s}_{23}+\Psi^{\rm s}_{31}\right)}\ris\,,
\end{eqnarray}
so that the holonomy accumulated by the symmetric Poincar\'e spinor around the closed loop is
the closed-loop Pancharatnam phase
\begin{equation}
\Psi^{\rm s}_{1231}= \Psi^{\rm s}_{12}+\Psi^{\rm s}_{23}+\Psi^{\rm s}_{31}  \pmod{2\pi}\,.
\end{equation}
Pancharatnam determined that this accumulated holonomy
equals one-half of the oriented solid angle 
\begin{equation}
\Omega_{1231}=2 \tan^{-1}\frac{\hat{\#r}_1\.\left(\hat{\#r}_2\times\hat{\#r}_3\right)}
{1+\hat{\#r}_1\.\hat{\#r}_2+\hat{\#r}_2\.\hat{\#r}_3+\hat{\#r}_1\.\hat{\#r}_3}
\end{equation}
subtended   at the center of the Poincar\'e sphere by the
spherical triangle ${\mathcal T}_{123}$ created by the geodesics joining location $1$ to $2$, $2$ to
$3$, and $3$ to $1$, i.e.,
\begin{equation}
\label{def-Omega1231-1}
\Psi^{\rm s}_{1231}=\displaystyle{\frac{\Omega_{1231}}{2}} \pmod{2\pi}\,.
\end{equation}
Since $\Psi^{\rm s}_{mn}=-\Psi^{\rm s}_{nm}$ according to Eq. (\ref{GPS2}), Eqs.~(\ref{def-123}) and (\ref{def-Omega1231-1}) yield
\begin{equation}
\Psi^{\rm s}_{123}-\Psi^{\rm s}_{13}=\displaystyle{\frac{\Omega_{1231}}{2}}  \pmod{2\pi}\,.
\label{eq24}
\end{equation}
Incidentally,   $\Psi^{\rm s}_{1231}$ is equal to the Berry phase for adiabatic systems \cite{Berry1987}.

\section{Spinors galore}\label{sec:sg}
Instead of  $\bpsis$ of  Eq.~(\ref{def-PS-0}), Das et al. \cite{Das} used $\bpsia$  of Eq.~(\ref{def-PA-0})
to define the geometric phase
\begin{equation}
\label{GPA2-def}
\Psia_{mn}={\rm Arg}\lec{\bpsiam}^\dag\.{\bpsian}\ric\,,
\end{equation}
where
\begin{equation}
\label{def-PA}
\bpsiam=\les
\begin{array}{c}
\cos\left(\frac{\pi}{4}-\frac{\beta_m}{2}\right)
\\[5pt]
\sin\left(\frac{\pi}{4}-\frac{\beta_m}{2}\right)\exp\left(i\alpha_m\right)
\end{array}
\ris= \exp\left(i\alpha_m/2\right)\,\bpsism
\,
\end{equation}
and $\bpsian$ is defined similarly.
Whereas $\Psis_{mn}$ is the \textsf{symmetric geometric phase}, 
$\Psia_{mn}$  
is the  \textsf{asymmetric geometric phase}.

The symmetric and the asymmetric Poincar\'e spinors are not the only options. Many others can be defined \cite{LakhtakiaAthens}. 
This paper is focused
on the  double-indexed Poincar\'e spinor  (DIPS)
\begin{equation}
\label{def-Ppq-0}
\bpsipq=
\les
\begin{array}{c}
\cos\les{p\left(\frac{\pi}{4}-\frac{\beta}{2}\right)}\ris\, \exp\les{i(q-1)\alpha/2}\ris
\\[5pt]
\sin\les{p\left(\frac{\pi}{4}-\frac{\beta}{2}\right)}\ris\, \exp\les{i(q+1)\alpha/2}\ris
\end{array}
\ris\,,\,\, p\in\mathbb{R}\,,\,\, q\in\mathbb{R}\,.
\end{equation}
Both  symmetric and  asymmetric Poincar\'e spinors are included in $\bpsipq$,
because
\begin{equation}
\bpsis= \les\underline{\tpsi}^{(1,0)}\ris\qquad{\rm and}\qquad
\bpsia= \les\underline{\tpsi}^{(1,1)}\ris\,.
\end{equation}

The DIPS can be written as
\begin{equation}
\bpsipq=\exp(iq\alpha/2)\exp\lec-i\alpha[\boldsymbol{\sigma}_z]/2\ric\.\exp\lec-i p\left(\frac{\pi}{2}-{\beta}\right)[\boldsymbol{\sigma}_y]/2\ric\.
\les\begin{array}{c}1\\0\end{array}\ris\,.
\end{equation}
Hence, $[\#U]\.\bpsipq$ is     also a valid spinor in the same SU(2) projective class with the same $(p,q)$ parametrization,
and $\bpsipq$ is SU(2)-covariant, if $p\ne0$.

{\bf Lemma~1.} \textit{
For any $[\#U]$,  the rotated spinor
 $[\#U]\.\bpsipq$ is again representable in the same DIPS form with the same $p\ne0$ and $q$,
up to an overall phase factor. In particular, there exist real parameters
$\alpha'$, $\beta'$, and $\chi$ such that}
\begin{equation}
[\#U]\.\bpsipq
=
\exp(i\chi)\les
\begin{array}{l}
\cos\left[p\left(\frac{\pi}{4}-\frac{\beta'}{2}\right)\right] \exp[{i(q-1)\alpha'/2}]\\[5pt]
\sin\left[p\left(\frac{\pi}{4}-\frac{\beta'}{2}\right)\right] \exp[{i(q+1)\alpha'/2}]
\end{array}\ris\,.
\label{eq-Ubsipq}
\end{equation}

\noindent{\bf Proof.}
Since $[\#U]$ is unitary, we have
\begin{equation}
[\#U]\.\bpsipq= \les \begin{array}{c}a_1\,\exp(i\varphi_1)\\ {a_2}\exp(i\varphi_2)\end{array}\ris\,,
\qquad a_1^2+a_2^2=1\,,
\end{equation}
with $a_1\in\mathbb{R}$ and $a_2\in\mathbb{R}$.
Let us define
\begin{equation}
\alpha'=\varphi_2-\varphi_1\,,
\qquad
\vartheta=2\tan^{-1}\left(\frac{a_2}{a_1}\right)\,,
\qquad
\beta'=\frac{\pi}{2}-\frac{\vartheta}{p}
\quad (p\neq 0),
\end{equation}
and choose the overall phase
\begin{equation}
\chi=\varphi_1-\frac{q-1}{2}\alpha'\,.
\end{equation}
Then,
\begin{subequations}
\begin{equation}
a_1\exp({-i\chi})
=
\cos(\vartheta/2)\,\exp\les{i(q-1)\alpha'/2}\ris\,,
\end{equation}
\begin{equation}
a_2\exp({-i\chi})
=
\sin(\vartheta/2)\,\exp\les{i(q+1)\alpha'/2}\ris\,,
\end{equation}
\end{subequations}
and Eq.~\eqref{eq-Ubsipq} follows.
Hence, the rotated spinor can again be written in DIPS form with the same $p$ and $q$,
up to a overall phase factor.

{\bf Lemma~2.} 
\textit{The DIPS $\bpsizeroq$ is not SU(2)-covariant as a family. Instead, it is covariant only under the
\(U(1)\subset SU(2)\) subgroup that preserves the north pole of the Poincar\'e sphere.}

\noindent{\bf Proof.}
\begin{equation}
[\#U]\.\bpsizeroq=\exp[{i(q-1)\alpha/2}] \begin{bmatrix} u_1\\-u_2^\ast\end{bmatrix}
\end{equation}
can be of the same form as $\bpsizeroq$ only if $u_2=0$. In that case, $u_1=\exp(i\delta)$ with $\delta\in\mathbb{R}$ and  $[\#U]$ belongs to subgroup
U(1) of SU(2) corresponding to rotations about the $z$ axis. Hence, $\bpsizeroq$ is not SU(2)-covariant,
but only U(1)-covariant.

\section{Double-indexed geometric phase}\label{sec:DIGP}
The definition of the DIPS is accompanied by that of the double-indexed overlap (DIO)
\begin{equation}
\label{def-DIO}
\tupsilon^{(p,q)}_{mn}= {\bpsipqm}^\dag\.{\bpsipqn}
\end{equation}
and the DIGP
\begin{equation}
\label{def-DIGP}
\Psipqmn=\Arg\lec\tupsilon^{(p,q)}_{mn}\ric  \,,
\end{equation}
with location $m$ serving as the reference location.
Note that $\Psipqmn$ is undefined when   the locations $m$ and $n$ are antipodal.

The special value 
\begin{equation}
\tPsi^{(0,1)}_{mn}=0
\end{equation}
emerges from the identity
\begin{equation}
\tPsi^{(0,q)}_{mn}=\frac{1-q}{2}\left(\alpha_m-\alpha_n\right)\pmod{2\pi}\,.
\end{equation}
The overlap of a DIPS with itself is unity, i.e.,
\begin{subequations}
\begin{equation}
\tupsilon^{(p,q)}_{mm}=1\,,
\end{equation}
leading to
\begin{equation}
\Psipqmm=0\,.
\end{equation}
\end{subequations}

Since   inner products reverse under complex conjugation, i.e.,
\begin{subequations}
\begin{equation}
\tupsilon^{(p,q)}_{mn}=\left(\tupsilon^{(p,q)}_{nm}\right)^\ast
\,,
\end{equation}
we get the geodesic-sense-reversal symmetry
\begin{equation}
\Psipqnm=-\Psipqmn \pmod{2\pi}\,.
\end{equation}
\end{subequations}

\subsection{Symmetries}\label{sec:symm}

As the DIPS is defined in terms of trigonometric functions of two geographic coordinates on the Poincar\'e sphere,
it is bound to deliver certain symmetries to the DIGP with respect to the indexes $p$ and $q$. 
First, as the DIO
\begin{subequations}
\begin{equation}
\tupsilon^{(-p,q)}_{mn}=  \tupsilon^{(p,q)}_{mn}\,
\end{equation}
is not affected by the sign reversal of $p$,
we get
\begin{equation}
\tPsi^{(-p,q)}_{mn}= \tPsi^{(p,q)}_{mn}   \pmod{2\pi}\,.
\label{eq33b}
\end{equation}
\end{subequations}
Second, since
\begin{equation}
\tupsilon^{(p,q)}_{mn}=\exp\les{-iq(\alpha_m-\alpha_n)/2}\ris \tupsilon^{(p,0)}_{mn}
\end{equation}
delivers
\begin{subequations}
\begin{equation}
\tupsilon^{(p,-q)}_{mn}=\exp\les{iq(\alpha_m-\alpha_n)}\ris \tupsilon^{(p,q)}_{mn}\,,
\end{equation}
we get
\begin{equation}
\tPsi^{(p,-q)}_{mn}= \tPsi^{(p,q)}_{mn} +q(\alpha_m-\alpha_n) \pmod{2\pi}\,
\label{eq35b}
\end{equation}
\end{subequations}
for the sign reversal of $q$.
From Eqs.~(\ref{eq33b}) and (\ref{eq35b}), it follows that
\begin{equation}
\tPsi^{(-p,-q)}_{mn}= \tPsi^{(p,q)}_{mn} +q(\alpha_m-\alpha_n)  \pmod{2\pi}\,,
\label{eq36}
\end{equation}
when the signs of  $p$ and $q$ are simultaneously reversed.

The effect of a shift $\delta_p$ in $p$ on ${\tPsi}_{mn}^{(p,q)}$ is  complicated, with
\begin{equation}
{\tPsi}_{mn}^{(p+\delta_p,q )} - {\tPsi}_{mn}^{(p,q )} =\Arg\les \dfrac{\tupsilon_{mn}^{(p+\delta_p,q)}}{\tupsilon_{mn}^{(p,q)}}\ris
\end{equation}
leading to
\begin{equation}
\dfrac{\partial}{\partial p} {\tPsi}_{mn}^{(p,q)}= {\Im}\les\dfrac{1}{{\tupsilon}_{mn}^{(p,q)}}\,
\dfrac{\partial}{\partial p} {\tupsilon}_{mn}^{(p,q)}\ris\,,
\end{equation}
where
\begin{eqnarray}
\nonumber
\dfrac{\partial}{\partial p} {\tupsilon}_{mn}^{(p,q)}
&=&-
\exp\left[-\,i\,q\,\frac{\alpha_m-\alpha_n}{2}\right]\times
\\ [5pt]
&&
\nonumber
\lec
\frac{\beta_m-\beta_n}{2}\,
\cos\left(\frac{\alpha_m-\alpha_n}{2}\right)
\sin\les{p}\frac{\beta_m-\beta_n}{2}\ris
\right.
\\[5pt]
&&
+\left.
i\left(\frac{\pi}{2}-\frac{\beta_m+\beta_n}{2}\right)
\sin\left(\frac{\alpha_m-\alpha_n}{2}\right)
\sin\les\frac{p}{2}\left(\pi-\beta_m-\beta_n\right)\ris
\ric\,.
\end{eqnarray}
There is no universal linear or multiplicative simplification for either ${\tupsilon}_{mn}^{(p,q)}$ or ${\tPsi}_{mn}^{(p,q)}$
 when $p \to p + \delta_p$.

In contrast, a shift of $\delta_q$ in $q$ contributes only a  phase factor to ${\tupsilon}_{mn}^{(p,q)}$ because
\begin{equation}
{\tupsilon}_{mn}^{(p,q+\delta_q)}=\exp\les{-i\delta_q\dfrac{\alpha_m-\alpha_n}{2}} \ris {\tupsilon}_{mn}^{(p,q)}\,.
\end{equation}
There is a consequential linear shift
in the DIGP ${\tPsi}_{mn}^{(p,q)}$ to
\begin{equation}
\label{eq41}
{\tPsi}_{mn}^{(p,q+\delta_q)}={\tPsi}_{mn}^{(p,q)}-\delta_q\dfrac{\alpha_m-\alpha_n}{2} \pmod{2\pi}\,,
\end{equation}
which leads to the partial derivative
\begin{equation}
\dfrac{\partial}{\partial q} {\tPsi}_{mn}^{(p,q)}=-\dfrac{\alpha_m-\alpha_n}{2} \pmod{2\pi}\,.
\end{equation}
The periodicity of ${\tPsi}_{mn}^{(p,q)}$ with respect to $q$ is implicit by virtue of Eq.~(\ref{eq41}), the period depending on
the longitudinal difference ${\alpha_m-\alpha_n}$.

\subsection{Closed loops}
Just like the closed-loop sum $\Psi^{s}_{1231}$ was interpreted by Pancharatnam to be equal to one-half of the solid angle subtended at the center of the
Poincar\'e sphere by the spherical triangle ${\mathcal T}_{123}$, its counterpart
\begin{equation}
\tPsi^{\left(p,q\right)}_{1231}=
\tPsi^{\left(p,q\right)}_{12}+\tPsi^{\left(p,q\right)}_{23}+\tPsi^{\left(p,q\right)}_{31}  \pmod{2\pi}
\label{eq43}
\end{equation}
can also be similarly interpreted---but after a mapping
of $(\alpha_m,\beta_m)$ to $(\alpha'_m,\beta'_m)$, $m\in\lec 1, 2,3\ric$,  as follows.
In terms of the unfolded latitude
\begin{equation}
\bar\beta_m' = p\beta_m + (1-p)\frac{\pi}{2}\,,
\label{eq44}
\end{equation}
 the DIPS satisfies the relation
\begin{equation}
\big[\tilde\psi^{(p,q)}(\alpha_m,\beta_m)\big]
= e^{iq\alpha_m/2}\,\big[\psi^{\mathrm s}(\alpha_m,\bar\beta_m')\big]\,.
\label{eq45}
\end{equation}
The equality in Eq.~(\ref{eq36}) is exact prior to reduction to
the canonical longitude $\alpha'_m\in[0,2\pi)$  and canonical latitude $\beta'_m\in[-\pi/2,\pi/2]$.
These canonical coordinates are obtained from
\begin{equation}
\alpha_m'=
\begin{cases}
\alpha_m \pmod{2\pi}, & \cos\bar\beta_m'\geq 0\,,
\\[5pt]
\alpha_m+\pi \pmod{2\pi}, & \cos\bar\beta_m'<0\,,
\end{cases}
\label{eq46}
\end{equation}
and
\begin{equation}
\beta_m' = \sin^{-1}(\sin\bar\beta_m')\,.
\label{eq47}
\end{equation}
Accordingly, the mapped locations are represented projectively by
\begin{equation}
\big[\tilde\psi^{(p,q)}(\alpha_m,\beta_m)\big]
\sim e^{iq\alpha_m/2}\,\big[\psi^{\rm s}(\alpha_m',\beta_m')\big],
\label{eq48}
\end{equation}
where $\sim$ denotes equality up to an overall phase factor.

The locations $1'$, $2'$, and $3'$---corresponding, respectively, to $(\alpha'_1,\beta'_1)$, $(\alpha'_2,\beta'_2)$,
and $(\alpha'_3,\beta'_3)$---form a spherical triangle ${\mathcal T}'_{123}$ that subtends the solid angle
\begin{equation}
\Omega'_{1231}=2 \tan^{-1}\frac{\hat{\#r}'_1\.\left(\hat{\#r}'_2\times\hat{\#r}'_3\right)}
{1+\hat{\#r}'_1\.\hat{\#r}'_2+\hat{\#r}'_2\.\hat{\#r}'_3+\hat{\#r}'_1\.\hat{\#r}'_3}
\end{equation}
at the center of the Poincar\'e sphere, where
\begin{equation}
\hat{\#r}'_m= \left(\hat{\#u}_1\cos\alpha'_m+\hat{\#u}_2\sin\alpha'_m\right)\cos\beta'_m +\hat{\#u}_3\sin\beta'_m\,.
\end{equation}
Then,
\begin{equation}
\tPsi^{\left(p,q\right)}_{1231} = \frac{\Omega'_{1231}}{2} \pmod{2\pi}\,
\end{equation}
follows from Eq.~(\ref{def-Omega1231-1}).

The closed-loop sum $\tPsi^{\left(p,q\right)}_{1231}$ is independent of $q$ because
\begin{subequations}
\begin{equation}
\tPsi^{\left(p,q\right)}_{1231}=\tPsi^{\left(p,0\right)}_{1231}  \pmod{2\pi} \,\,\forall q\in\mathbb{R}\,.
\label{eq52a}
\end{equation}
Accordingly,
\begin{equation}
\dfrac{\partial}{\partial q} \tPsi^{\left(p,q\right)}_{1231}=0\,
\label{eq52b}
\end{equation}
and $\tPsi^{\left(p,q\right)}_{1231}$ is not affected by the shift $q\to q+\delta_q$.
\end{subequations}
Furthermore, although
\begin{equation}
\tPsi^{\left(1,q\right)}_{1231}=\Psi^{s}_{1231} \pmod{2\pi} \,\,\forall q\in\mathbb{R}\,,
\end{equation}
in general
\begin{equation}
\tPsi^{\left(p,q\right)}_{1231}\ne\Psi^{s}_{1231}\,,\,\,p\ne1\,.
\end{equation}
The effect of a shift of $p$ to $p+\delta_p$ on $\tPsi^{\left(p,q\right)}_{1231}$ is complicated, as can be surmised from
Sec.~\ref{sec:symm} and the relationship of $(\alpha_m',\beta_m')$ to $(\alpha_m,\beta_m)$, $m\in\lec1,2,3\ric$.

Equations~(\ref{eq33b}) and (\ref{eq43}) together yield
\begin{subequations}
\begin{equation}
\tPsi^{(-p,q)}_{1231}= \tPsi^{(p,q)}_{1231}   \pmod{2\pi}\,,
\label{eq55a}
\end{equation}
and Eq. (\ref{eq52a}) leads to
\begin{equation}
\tPsi^{(p,-q)}_{1231}= \tPsi^{(p,q)}_{1231}   \pmod{2\pi}\,.
\label{eq55b}
\end{equation}
From both foregoing symmetries, we get
\begin{equation}
\tPsi^{(-p,-q)}_{1231}= \tPsi^{(p,q)}_{1231}   \pmod{2\pi}\,.
\label{eq55c}
\end{equation}
\end{subequations}

\section{Examples}\label{sec:examples}
This section is devoted to two scattering problems for which the application of $\tPsi^{(p,q)}_{mn}$ is expected to be useful
in the coming decade. Instead of limiting the applicability of the geometric-phase concept to two plane waves propagating
in the same direction, the two examples illustrate the application of the DIGP concept to  two non-co-propagating plane waves.

Before continuing with the examples, note that  polarimetric measurements of the two plane waves located at $1$ and $2$ on the Poincar\'e
sphere are needed to calculate a collection of multiple  DIGPs $\tPsi^{(p,q)}_{12}$. In other words, determination of $\tPsi^{(p,q)}_{12}$ for a
range of values of $p$ and $q$ does not entail an experimental burden greater than to determine either $\Psis_{12}$ or $\Psia_{12}$.

\subsection{Plane-wave scattering by thin films}\label{sec:thin films}
When a plane wave is incident on a thin film that is piecewise homogeneous in the thickness direction,   a
specularly reflected plane wave and a specularly transmitted plane wave arise. Knowing the frequency, the polarization
state,  the propagation direction, and field amplitudes of the incident plane wave, the same quantities can be determined for the reflected
and transmitted plane waves by solving a two-point boundary-value problem. Several semi-analytical techniques exist to solve
that problem, of which the most versatile is
the 4$\times$4 transfer-matrix method \cite{TMMEO} that can accommodate linear bianisotropic
materials which are far more complicated than isotropic dielectric materials.

Two sets of quantities can be calculated thereafter from the reflection and transmission coefficients \cite{LakhCBPupdated}. The first set comprises
intensity-dependent quantities such as linear and circular reflectances and transmittances as well as linear and circular dichroisms. The 
second set comprises ellipticity, optical rotation, and the Pancharatnam phase of the reflected and transmitted plane waves relative
to the incident plane wave. The focus
of this paper is on expanding the second set to include the DIGPs of the reflected and transmitted plane waves with
 the incident plane wave serving as the reference.

\subsubsection{Incident, reflected, and transmitted plane waves}
A plane
wave, propagating in the half-space
$z < 0$ at an angle $\thetainc\in[0,\pi/2)$ with respect to the $+z$ axis and at an angle $\phiinc\in[0,2\pi)$ with respect
to the $x$ axis in the $xy$ plane, is incident on a planar thin film of thickness $L$. Analogously to the representation of ${\#E}^+$ in Eqs.~\eqref{eqE-lin}
and \eqref{eqE-circ},
the electric  field phasor associated
with the incident plane wave is represented as \cite{STFbook}
\begin{subequations}
\begin{eqnarray}
\Ei&=& \le \aas \spinc + \aap\pinc\ri \winc \exp{\left(i\ko z{\cos\thetainc}\right)}\,, \qquad   z < 0\,,
\label{eqEi-lin}
\\
\nonumber
&=&
\les \frac{\le i\spinc - \pinc \ri}{\sqrt{2}} \, \aal -\, \frac{\le i\spinc +\pinc
\ri}{\sqrt{2}} \, \aar \ris
\\
 &&\qquad \times\,\winc \exp{\left(i\ko z{\cos\thetainc}\right)}\,, \qquad   z < 0\,,
 \label{eqEi-circ}
\end{eqnarray}
\end{subequations}
where
 \begin{equation}
 \left.\begin{array}{l}
\winc =
\exp{\les i\ko \le x\cos\phiinc + y\sin\phiinc\ri\sin\thetainc\ris}\\[5pt]
\spinc=-\ux\sin\phiinc + \uy \cos\phiinc \\[5pt]
\pinc=-\le \ux \cos\phiinc + \uy \sin\phiinc \ri \cos{\thetainc} 
 + \uz \sin{\thetainc}
\end{array}
\right\}
\, .
\end{equation}
  Linear  incidence amplitudes are denoted by $\aas$ and $\aap$  in Eq.~(\ref{eqEi-lin}), whereas the
circular  incidence amplitudes are denoted by $\aal$ and $\aar$ in Eq.~(\ref{eqEi-circ}).

The   electric   field phasor of the specularly reflected plane wave is expressed as
\begin{subequations}
\begin{eqnarray}
\label{eqEr-lin}
\Er&=& \le \bbs\spinc+\bbp\pref\ri \winc \exp\left({-i\ko   z\cos{\thetainc}}\right) \,,\qquad z < 0\,,
\\
\nonumber
&=&-\, \les \frac{\le i\spinc - \pref \ri}{\sqrt{2}} \, {\bbl} -\, \frac{\le i\spinc +
\pref \ri}{\sqrt{2}} \, {\bbr} \ris  \\
&&\qquad\times\,\winc\exp\left({-i\ko   z\cos{\thetainc}}\right)\,,\qquad z < 0 \, ,
\label{eqEr-circ}
\end{eqnarray}
\end{subequations}
where
 \begin{equation}
 \pref=\le \ux \cos\phiinc + \uy \sin\phiinc \ri \cos{\thetainc} 
 + \uz \sin{\thetainc}\, .
\end{equation}
The electric   field phasor of the specularly transmitted plane wave is represented similarly to that of the incident plane wave as
\begin{subequations}
\begin{eqnarray}
\label{eqEt-lin}
\Et&=&\le \ccs\spinc +\ccp\pinc\ri   \winc
\exp\les{i\ko   (z-L)\cos{\thetainc}}\ris\,,\qquad z > L \,,
\\
\nonumber
&=& \les \frac{\le i\spinc - \pinc \ri}{\sqrt{2}} \, {\ccl} -\, \frac{\le i\spinc +\pinc
\ri}{\sqrt{2}} \, {\ccr} \ris \\
\label{eqEt-circ}
& &  \times \, \winc \exp\les{i\ko   (z-L)\cos{\thetainc}}\ris \, , \qquad z > L \, .
\end{eqnarray}
\end{subequations}
Linear reflection amplitudes are denoted by $\bbs$ and $\bbp$  in Eq.~(\ref{eqEr-lin}), whereas the
circular reflection amplitudes are denoted by $\bbl$ and $\bbr$ in Eq.~(\ref{eqEr-circ}).
Similarly,  $\ccs$ and $\ccp$ are the linear transmission amplitudes in Eq.~(\ref{eqEt-lin}),
whereas $\ccl$ and $\ccr$ are the circular transmission amplitudes in Eq.~(\ref{eqEt-circ}). Note that
$\Er$ and $\Ei$ do not propagate in the same direction, although $\Et$ and $\Ei$ do.

\subsubsection{Boundary-value problem}
Provided that the thin film is piecewise homogeneous along the $z$ axis
and homogeneous along the $x$ and $y$ axes, the two-point boundary-value can be formulated as
the matrix equation
\begin{subequations}
\begin{equation}
\begin{bmatrix}
\ccs \\
\ccp \\
0 \\
0 \\
\end{bmatrix}
= [\#K_{\rm inc}]^{-1}\.[\#M] \. [\#K_{\rm inc}]\.
\begin{bmatrix}
\aas \\
\aap \\
\bbs \\
\bbp \\
\end{bmatrix} \,;
\label{4by4-lin}
\end{equation}
equivalently,
\begin{equation}
\begin{bmatrix}
i(\ccl-\ccr) \\
-(\ccl+\ccr)  \\
0 \\
0 \\
\end{bmatrix} 
= [\#K_{\rm inc}]^{-1}\. [\#M] \.[\#K_{\rm inc}]\.
\begin{bmatrix}
i(\aal-\aar) \\
-(\aal+\aar)  \\
-i(\bbl-\bbr) \\
\bbl+\bbr \\
\end{bmatrix} 
\,.
\label{4by4-circ}
\end{equation}
\end{subequations}
The 4$\times$4 matrix
\begin{equation}
[\#K_{\rm inc}] =
\begin{bmatrix}
-\sin \phiinc & -\cos\phiinc \cos\thetainc  & -\sin\phiinc & \cos\phiinc \cos\thetainc  \\
\cos\phiinc & -\sin\phiinc\cos\thetainc  & \cos\phiinc & \sin\phiinc\cos\thetainc  \\
-\etao^{-1} \cos\phiinc \cos\thetainc  &  \etao^{-1} \sin\phiinc & \etao^{-1} \cos\phiinc \cos\thetainc  & \etao^{-1} \sin\phiinc\\
-\etao^{-1} \sin\phiinc \cos\thetainc  & -\etao^{-1}\cos\phiinc & \etao^{-1} \sin\phiinc \cos\thetainc  & -\etao^{-1} \cos\phiinc\\
\end{bmatrix} \,
\end{equation}
contains the propagation direction of the incident plane wave along with $\etao=\sqrt{\muo/\epso}$ as the intrinsic
impedance of free space. 

The 4$\times$4 matrix $[\#M]$ captures the $z$-dependent and $\omega$-dependent
linear constitutive parameters of the thin film wholly filling up the region $0<z<L$ \cite{TMMEO,ESW}.
Several different ways have been devised to construct $[\#M]$ and then solve Eqs.~\eqref{4by4-lin} and \eqref{4by4-circ}
\cite{Herzinger,Polo2002,Berreman,Rokushima,oblique,pert,Venugopal,Oldano,Dreher}, of which
the piecewise uniform approximation technique \cite[Sec. 9.2.3]{STFbook} is perhaps the most convenient one. 

Suppose that the thin film is periodic with $N$ unit cells of thickness $L/N$ spanning the region $0<z<L$. In that case,
$[\#M] = [\#M_{\rm uc}]^N$, where the 4$\times$4 matrix $[\#M_{\rm uc}]$ holds for a  unit cell in isolation
from all the other  unit cells.

Suppose that a stack of three different thin films occupies the region $0<z<L$. Then, $[\#M] = [\#M_{3}]\.[\#M_{2}]\.[\#M_{1}]$,
where $[\#M_{\ell}]$ holds for the $\ell$th thin film in isolation,   the thin film labeled $\ell=1$ being the one that is illuminated
by the incident plane wave  \cite{STFbook}.

The solution of Eq.~\eqref{4by4-lin} or \eqref{4by4-circ} delivers the reflectance
\begin{equation}
R=\dfrac{\vert\bbl\vert^2+\vert\bbr\vert^2}{\vert\aal\vert^2+\vert\aar\vert^2}=
\dfrac{\vert\bbs\vert^2+\vert\bbp\vert^2}{\vert\aas\vert^2+\vert\aap\vert^2}
\end{equation}
and the transmittance
\begin{equation}
T=\dfrac{\vert\ccl\vert^2+\vert\ccr\vert^2}{\vert\aal\vert^2+\vert\aar\vert^2}=
\dfrac{\vert\ccs\vert^2+\vert\ccp\vert^2}{\vert\aas\vert^2+\vert\aap\vert^2}\,.
\end{equation}
Other remittances (e.g., $\vert\ccl\vert^2/\vert\aar\vert^2$ when $\aal=0$) are also calculated by researchers for
diverse purposes \cite{STFbook,BW1998}.

\subsubsection{DIGPs of reflected and transmitted plane waves}
The Stokes parameters of the incident plane wave are given by \cite{Jackson}
\begin{equation}
\label{Stokes-inc}
\left.\begin{array}{l}
\sinczero = \vert\aal\vert^2+\vert\aar\vert^2 =\vert\aas\vert^2+\vert\aap\vert^2
\\[5pt]
\sincone =2\,\Re\left(\aal\,\aar^\ast\right)=\vert\aap\vert^2-\vert\aas\vert^2
\\[5pt]
\sinctwo =2\,\Im\left(\aal\,\aar^\ast\right)=2\,\Re\left(\aas\,\aap^\ast\right)
\\[5pt]
\sincthree = \vert\aar\vert^2-\vert\aal\vert^2=2\,\Im\left(\aas\,\aap^\ast\right)
\end{array}
\right\}\,.
\end{equation} 
The longitude $\alphainc$ and the latitude $\betainc$ can then be calculated using Eqs.~\eqref{def-alphabeta}
and \eqref{Stokes-inc}, and the DIPS $\bpsipqinc$ of the incident plane wave thereafter  using
Eq.~\eqref{def-Ppq-0}.

After making the   changes $\lec\aal\to\ccl,\,\aar\to\ccr,\,\aas\to\ccs,\,\aap\to\ccp\ric$, Eqs.~(\ref{Stokes-inc}) 
can be used to determine the Stokes parameters $\strazero$,  $\straone$,  $\stratwo$, and  $\strathree$
of the transmitted plane wave, and the   DIPS $\bpsipqtra$ of the transmitted plane wave 
can then be obtained similarly.
Calculation of the DIPS $\bpsipqref$ of the reflected plane wave will follow the same route \cite{Lakh-ReuschGP}.
The DIGP of the reflected plane wave can then be defined as
\begin{equation}
\label{def-DIGP-ref}
\Psipqref=\Arg\lec{\bpsipqinc}^\dag\.{\bpsipqref}\ric  
\end{equation}
and the DIGP of the transmitted plane wave as
\begin{equation}
\label{def-DIGP-tra}
\Psipqtra=\Arg\lec{\bpsipqinc}^\dag\.{\bpsipqtra}\ric  \,,
\end{equation}
with respect to the incident plane wave, following Eq.~\eqref{def-DIGP}.

\subsubsection{Chiral sculptured thin film}\label{CSTF}

The relative permittivity dyadic of a chiral sculptured thin film (CSTF)
is given by \cite{STFbook,Erten2015,McAtee2018}
\begin{eqnarray}
 \nonumber
&&\=\eps_{\rm CSTF}(z)=
\=S_{\rm z}(h,D,z) \.
\=S_{\rm y}(\chi)\.\left[\epsa\uz\uz+\epsb\ux\ux
+\epsc\uy\uy\right]
\nonumber
\\[5pt]
&&\qquad 
\.\=S_{\rm y}^{-1}(\chi) \.\=S_{\rm z}^{-1}(h,D,z)\,,
\quad z\in(0,L)\,.
\label{epsCSTF}
\end{eqnarray}
 The frequency-dependent relative permittivity scalars
$\epsa$, $\epsb$, and $\epsc$ embody   local orthorhombicity \cite{Nye}.
 The inclination dyadic
\begin{equation}
\=S_{\rm y} (\chi) = \uy\uy + \left(\ux\ux+\uz\uz\right)\cos\chi
  + \left(\uz\ux-\ux\uz\right)\sin\chi
 \label{SY-def}
 \end{equation}
contains $\chi\in[{0,\pi/2}]$ as an angle of inclination with respect to the $xy$ plane.
The structural chirality of the CSTF is captured
by the rotation dyadic
\begin{equation}
\=S_z(h,D,z)=\uz\uz +\left(\ux\ux+\uy\uy\right)\cos\left(2h\frac{\pi z}{D}\right)
+\left(\uy\ux-\ux\uy\right)\sin\left(2h\frac{\pi z}{D}\right)\,,
\label{Sz-def}
\end{equation} 
where $D$ is the structural period in the thickness direction (i.e., along the $z$ axis), whereas
$h \in\left\{-1,1\right\}$ is the structural-chirality parameter, with $h=-1$ for   left-handedness and
 $h=1$ for   right-handedness. Furthermore, the non-homogeneity of the CSTF along the $z$ axis  is encapsulated
 in $\=S_z(h,D,z)$.

The foregoing equations apply also to
 chiral smectic liquid crystals \cite{Parodi,Garoff,Abdulhalim},
with $\epsa\ne\epsb\ne\epsc$ and $\chi>0$. In addition, they apply to cholesteric liquid crystals with $\epsa\ne\epsb=\epsc$ and $\chi=0$
\cite{deG,Nityananda} and heliconical cholesteric liquid crystals \cite{Xiang} with $\epsa\ne\epsb=\epsc$ and $\chi\in(0,\pi/2)$.
The matrix $[\#M]$ for  CSTFs is available elsewhere \cite{STFbook}.

The hallmark of a CSTF is the circular Bragg phenomenon,
which is the circular-polarization-state-selective reflection of plane waves in a spectral regime called
the circular Bragg regime that depends on the direction of incidence \cite{LakhCBPupdated,LVaeu,FLaop}.  The reflectance
is very high when LCP light is incident on a structurally left-handed CSTF, provided that (i)  the direction of incidence
is not too oblique with respect to the thickness direction, (ii)
$\lambdao$ lies in the circular Bragg regime,  and (iii) the number of periods in CSTF is sufficiently large; however,
when RCP light is incident on a structuraly left-handed CSTF, the reflectance is very low in the circular
Bragg regime.
An analogous statement holds for structurally right-handed CSTFs. The circular Bragg phenomenon is resilient against structural disorder \cite{dcs}
and the tilt of the axis of  periodicity \cite{Slant1,Slant2}.

\begin{figure}[!htb]
\centering
 \includegraphics[width=6.5cm]{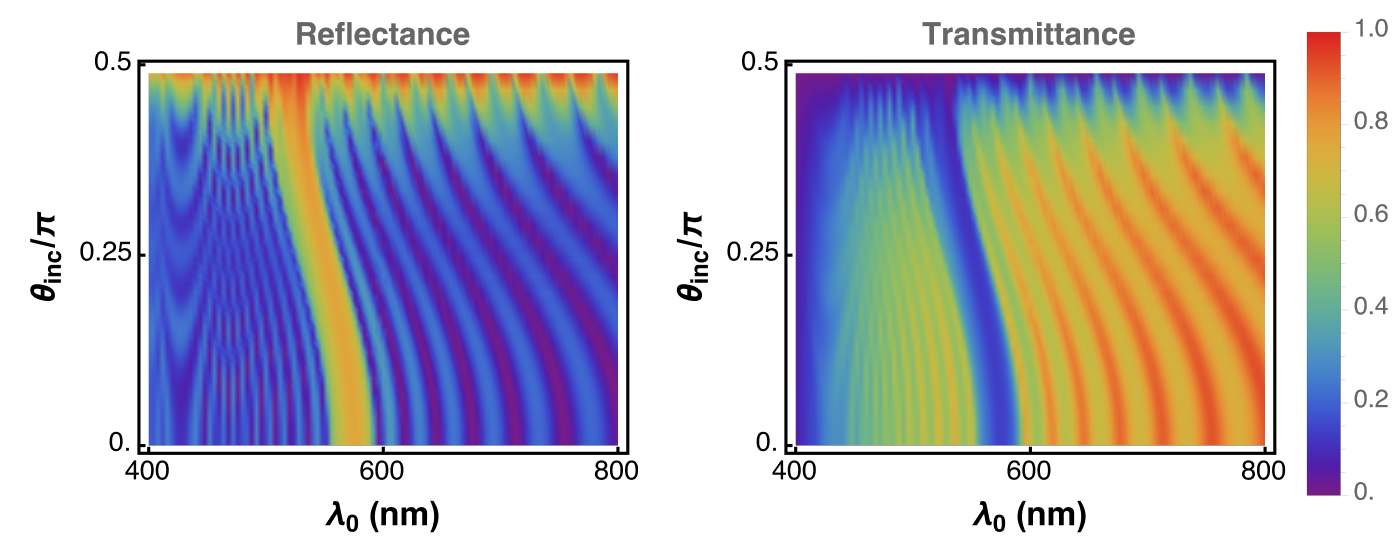} 
 \caption{\label{CSTF-remit} 
 Reflectance $R$ and transmittance $T$ as functions of
$\lambdao\in[400,800]$~nm and $\thetainc\in[0,\pi/2)$ when $\phiinc=\pi/4$, $\aas=1 + 0.3 i$, and $\aap=0.7 - 1.2 i$, for a 10-period-thick, structurally right-handed CSTF with constitutive parameters specified in Sec.~\ref{CSTF}.  
}
\end{figure}

Figure~\ref{CSTF-remit} provides  the reflectance $R$ and transmittance $T$   
as  functions of $\lambdao\in[400,800]$~nm and $\thetainc\in[0,\pi/2)$ when $\phiinc=45\deg$, $\aas=1 + 0.3 i$, and $\aap=0.7 - 1.2 i$,  calculated
for a 10-period-thick CSTF with the relative permittivity dyadic specified by Eqs.~\eqref{epsCSTF}--\eqref{Sz-def}. The relative permittivity scalars
\begin{equation}
\label{resonance}
\eps_{\ell}(\lambdao) = 1+ \frac{p_{\ell}}
{1 + (N_{\ell}^{-1}  - i\lambdao^{-1} \lambda_{\ell}  )^2}\,,\quad \ell\in\lec{\rm a, b, c}\ric\,,
\end{equation}
are single-resonance Lorentzian functions of the free-space wavelength $\lambdao$
\cite{Wooten},
this choice being consistent with the requirement of causality 
\cite{Frisch,Silva,WL96,Kinsler-EJP}.
The oscillator strengths are determined by the values of $p_{\ell}$,  $\lambda_{\ell} (1 + N_{\ell}^{-2})^{-1/2} $  are the  resonance wavelengths, and $\lambda_{\ell}/N_{\ell}$ are the  resonance linewidths, $\ell\in\left\{{\rm a, b, c}\right\}$.
 Values of the parameters used for all theoretical results reported in Secs.~\ref{CSTF} and \ref{CSTF-twist} are as follows:  $p_{\rm a} = 1.024$, $p_{\rm b} =1.990$, $p_{\rm c} =1.458 $, $\lambda_{\rm a} = \lambda_{\rm b}=\lambda_{\rm c} =360$~nm,   
$N_{\rm a} = N_{\rm b} =N_{\rm c}=500$, $\chi = 59\deg$, $h=1$, $L=10D$,  and $D = 320$~nm.   Data were generated on
a grid of 101 regularly spaced values of $\lambdao$
and 45 regularly spaced values of $\thetainc$. All calculations were performed 
on Mathematica\texttrademark~version~14.0.0.0.

The incident plane wave is elliptically polarized, having both LCP and RCP components. Its RCP component activates the circular Bragg phenomenon.
This is the reason for the presence of a high-intensity ridge in the map of $R$ and the low-intensity trough in the map of $T$ in Fig.~\ref{CSTF-remit}.
The circular Bragg regime blueshifts as the angle of incidence $\thetainc$ increases.
In addition, Fabry--P\'{e}rot fringes  present on the long-wavelength side of the circular Bragg regime arise from the finite thickness
of the CSTF.

%%%%%%%%%%%%%%%%%% Figure 1 begins %%%%%%%%%%%%%%%%%%
%%%%%%%%%%%%%%%%%% Figure 1 ends %%%%%%%%%%%%%%%%%%

\begin{figure}[!htb]
\centering
 \includegraphics[width=9.5cm]{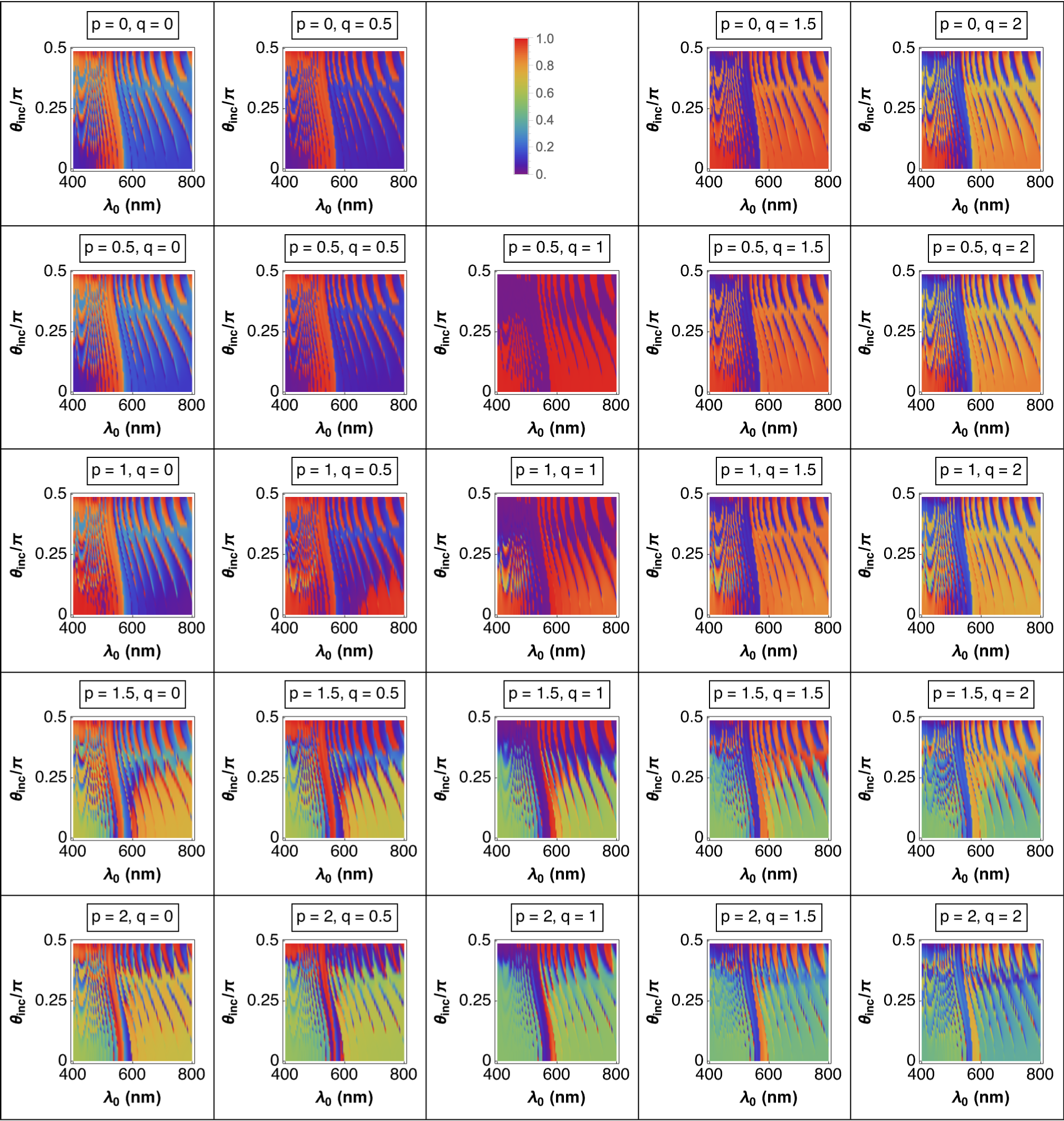} 
 \caption{\label{CSTF-ref} 
 $\Psipqref$ as a function of  $\lambdao\in[400,800]$~nm and $\thetainc\in[0,\pi/2)$ when $\phiinc=\pi/4$, $\aas=1 + 0.3 i$, and $\aap=0.7 - 1.2 i$, for a 10-period-thick, structurally right-handed CSTF with constitutive parameters specified in Sec.~\ref{CSTF}. Individual maps depict   $\Psipqref$  for $p\in\lec 0,0.5,1,1.5,2\ric$
 and $q\in\lec 0,0.5,1,1.5,2\ric$. Note that $\tPsi^{(0,1)}_{\rm ref}\equiv 0$.
}
\end{figure}

\begin{figure}[!htb]
\centering
 \includegraphics[width=9.5cm]{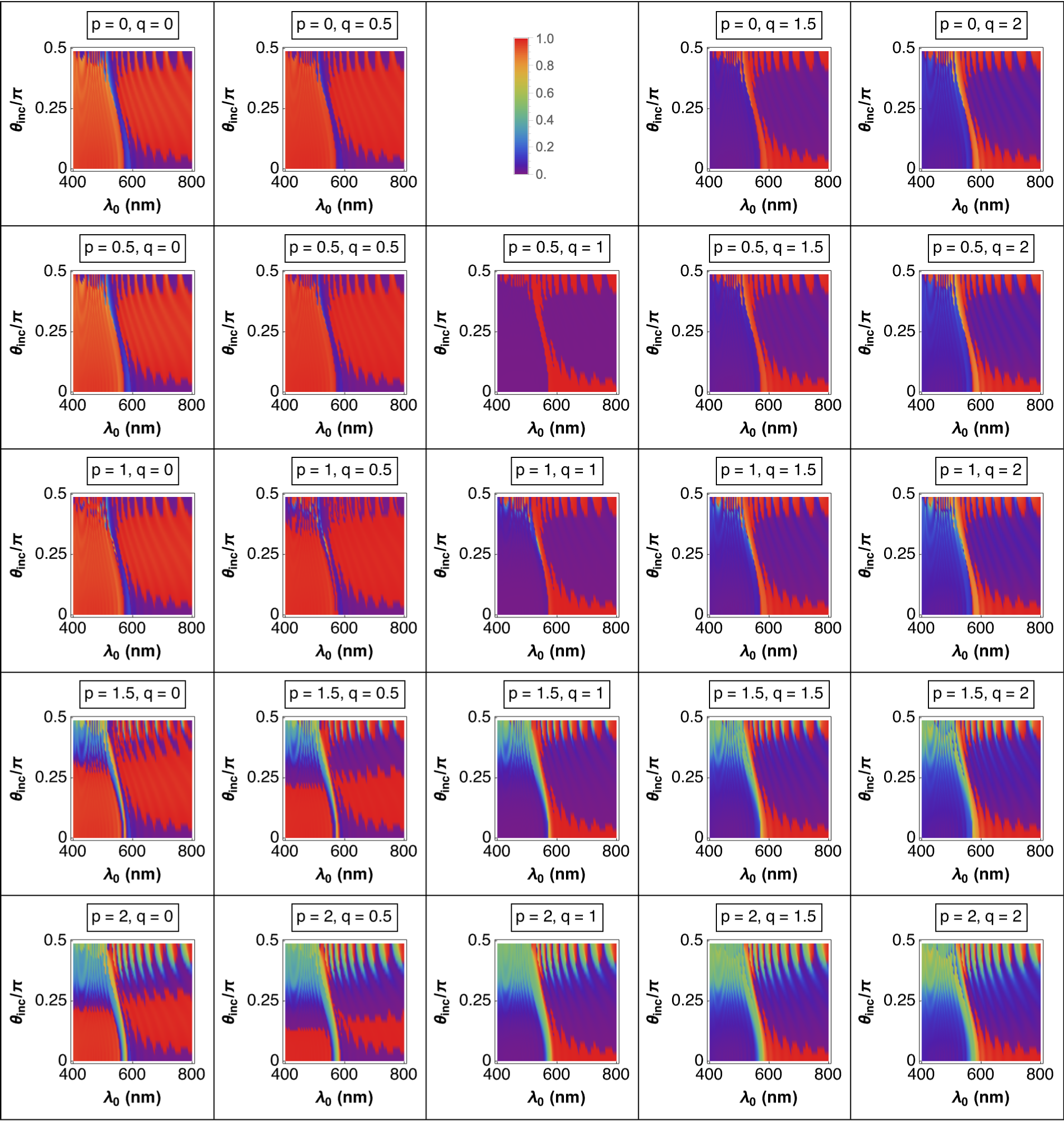} 
 \caption{\label{CSTF-tra} 
 $\Psipqtra$ as a function of  $\lambdao\in[400,800]$~nm and $\thetainc\in[0,\pi/2)$  when $\phiinc=\pi/4$, $\aas=1 + 0.3 i$, and $\aap=0.7 - 1.2 i$, for a 10-period-thick,
 structurally right-handed CSTF with constitutive parameters specified in Sec.~\ref{CSTF}. Individual maps depict   $\Psipqtra$  for $p\in\lec 0,0.5,1,1.5,2\ric$
 and $q\in\lec 0,0.5,1,1.5,2\ric$. Note that $\tPsi^{(0,1)}_{\rm tr}\equiv 0$.
}
\end{figure}

Figure~\ref{CSTF-ref} provides $\Psipqref$ and Fig.~\ref{CSTF-tra} provides  $\Psipqtra$ calculated
as  functions of $\lambdao\in[400,800]$~nm and $\thetainc\in[0,90\deg)$    for the same boundary-value problem \cite{Lakh2024josab,Das,Mondal}.
Neither $\Psipqref$ nor $\Psipqtra$ varies monotonically with $\lambdao$  for fixed $\thetainc$, or with
 $\thetainc$ for fixed $\lambdao$, except, trivially,
when $p=0$ and $q=1$. Even a casual visual comparison of Fig.~\ref{CSTF-remit} with Figs.~\ref{CSTF-ref} and \ref{CSTF-tra} is sufficient
to convince the reader that the circular Bragg phenomenon is evident in the DIGP maps.
The steepest  gradients of $\Psipqref$ and $\Psipqtra$ occur in the circular Bragg regime.  Away from this regime, both
$\Psipqref$ and $\Psipqtra$ vary smoothly, consistent with weak interaction. The parameters $p$ and $q$   modulate the
magnitudes of $\Psipqref$ and $\Psipqtra$, without obscuring the  circular Bragg phenomenon. Thus, the DIGP maps encode certain
aspects of the handedness-selective scattering physics.

%%%%%%%%%%%%%%%%%% Figure 2 begins %%%%%%%%%%%%%%%%%%
%%%%%%%%%%%%%%%%%% Figure 2 ends %%%%%%%%%%%%%%%%%%
%%%%%%%%%%%%%%%%%% Figure 3 begins %%%%%%%%%%%%%%%%%%
%%%%%%%%%%%%%%%%%% Figure 3 ends %%%%%%%%%%%%%%%%%%

Figures~\ref{CSTF-ref} and \ref{CSTF-tra} depict the systematic dependences of $\Psipqref$ and
$\Psipqtra$ on both indexes $p$ and $q$. Whereas $q$  produces an essentially linear  offset on the equator of the Poincar\'e sphere, 
$p$ reshapes the geometric-phase landscape by altering the relation between the  polarization state and the effective latitude on the Poincar\'e sphere. Thus, 
the DIGP maps for different $(p,q)$ pairs provide different spinor gauges or projections of the same scattering process, each emphasizing distinct phase contrasts. The family of DIGP maps in Fig.~\ref{CSTF-ref} can be viewed as a multi-parameter phase tomography of the CSTF reflection response, and the
family of DIGP maps in Fig.~\ref{CSTF-tra} can be viewed similarly for CSTF transmission response. 

Note that for any fixed $(p,q)$, not all features in the
map of $\Psipqref$  have counterparts in the map of $\Psipqtra$. Thus, $\Psipqref$  and $\Psipqtra$ should provide somewhat different information
about the structural chirality,   angle of inclination, dielectric anisotropy, and resonance structure of the CSTF. 

In principle, a collection of maps of $\Psipqref$  and $\Psipqtra$  for several 
$(p,q)$ pairs could be useful in uncovering the morphological and electromagnetic characteristics of CSTFs. A natural analytical strategy would be to treat the DIGPs as features in an inverse-scattering problem: quantify variations in DIGP maps to CSTF parameters,
build a forward model from  
 calculations \cite{STFbook,TMMEO}, and then fit measured DIGP data using least-squares or Bayesian inference \cite{Tarantola,Kaipio,Hansen}. A reduced-order decomposition, such as principal-component or singular-value analysis \cite{Hansen,Golub,Jolliffe}, could then identify which $(p,q)$ pairs are most 
 informative and whether an ensemble of DIGP maps is sufficient for robust parameter retrieval.

\subsubsection{Chiral sculptured thin film with central $90\deg$-twist defect}\label{CSTF-twist}

\begin{figure}[!htb]
\centering
 \includegraphics[width=6.5cm]{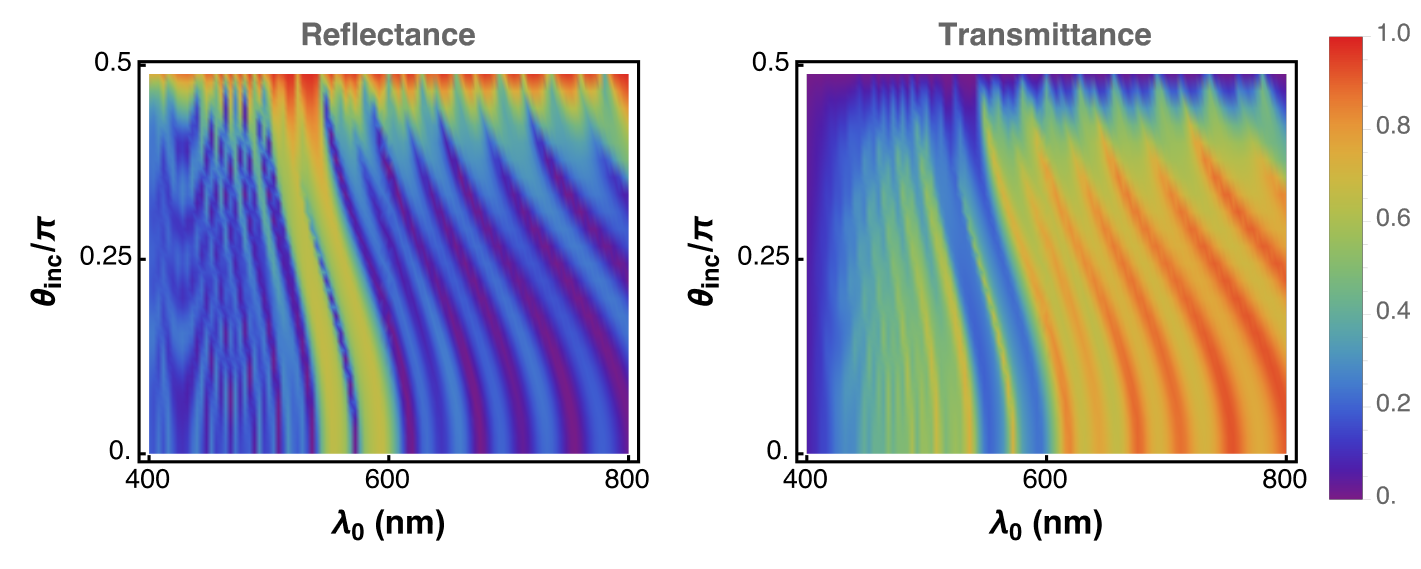} 
 \caption{\label{CSTF-twist-remit} 
Same as Fig.~\ref{CSTF-remit} except that the set of five unit cells in the back of the CSTF are twisted by $\pi/2$ about the $+z$ axis.
}
\end{figure}

When one-half of a CSTF with an even number of periods is rotated by $90\deg$ about the $z$ axis, the circular Bragg phenomenon is pierced
by a spectral reflection hole if the number of periods is relatively small \cite{Spacerless,Crossover1,Crossover2,Lakh2024pra}. This feature
is so distinct that it can be used for optical sensing applications \cite{Sensing}. It is exemplified by the maps
of $R$ and $T$ in the $\lambdao$--$\thetainc$ plane drawn in Fig.~\ref{CSTF-twist-remit} for the same CSTF as in Sec.~\ref{CSTF} except 
that the set of five unit cells in the back of the CSTF are twisted by $90\deg$ about the $+z$ axis. Whereas the circular Bragg regime
appears as  a $\sim$40-nm-wide feature for fixed $\thetainc$  in Fig.~\ref{CSTF-remit}, it shows up bifurcated  by a  
$\sim$10-nm-wide feature in Fig.~\ref{CSTF-twist-remit}.   Fabry--P\'{e}rot fringes  remain on the long-wavelength side of the circular Bragg regime.

%%%%%%%%%%%%%%%%%% Figure 4 begins %%%%%%%%%%%%%%%%%%
%%%%%%%%%%%%%%%%%% Figure 4 ends %%%%%%%%%%%%%%%%%%

\begin{figure}[!htb]
\centering
 \includegraphics[width=9.5cm]{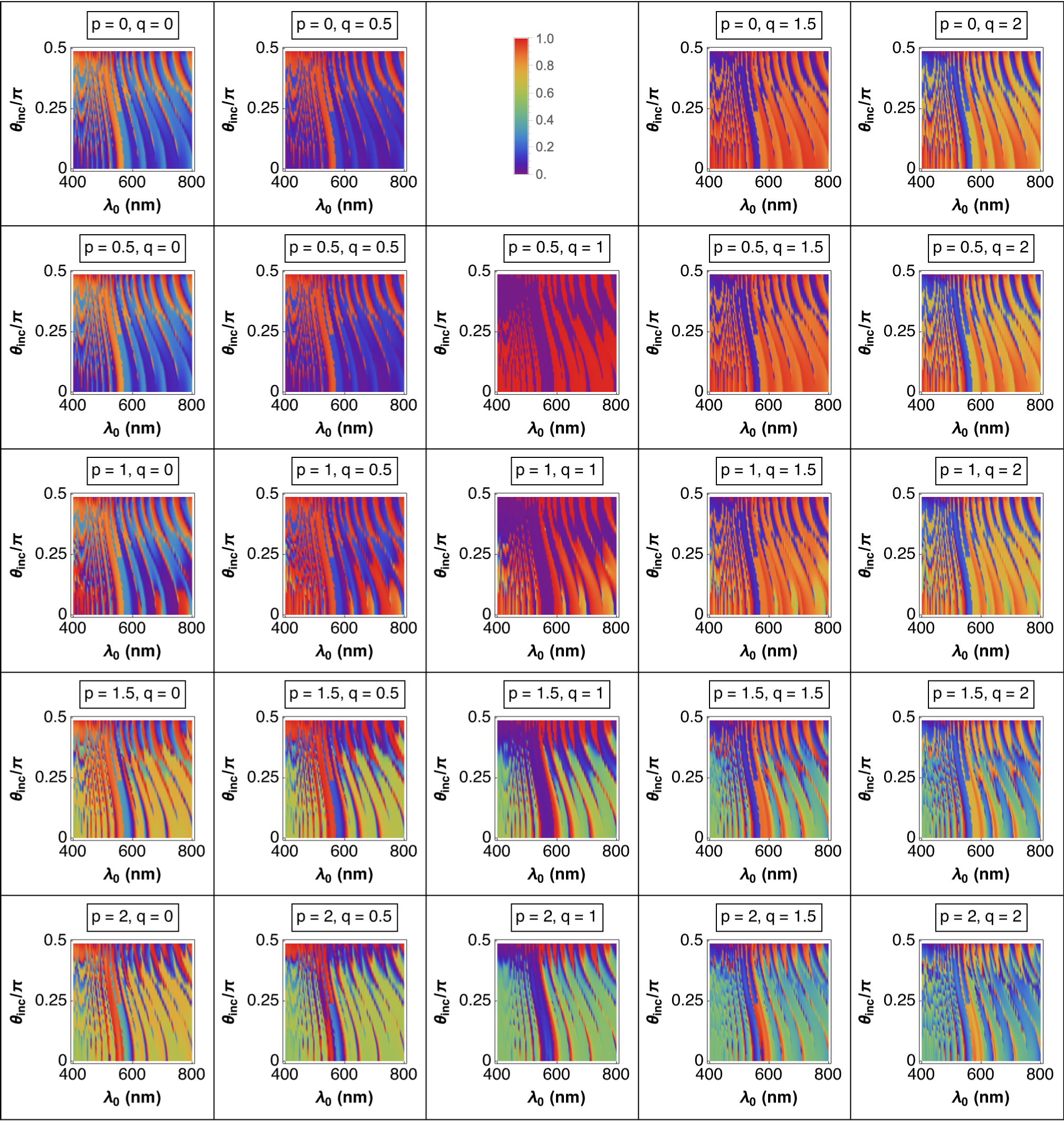} 
 \caption{\label{CSTF-twist-ref} 
Same as Fig.~\ref{CSTF-ref} except that the set of five unit cells in the back of the CSTF are twisted by $\pi/2$ about the $+z$ axis.
}
\end{figure}

\begin{figure}[!htb]
\centering
 \includegraphics[width=9.5cm]{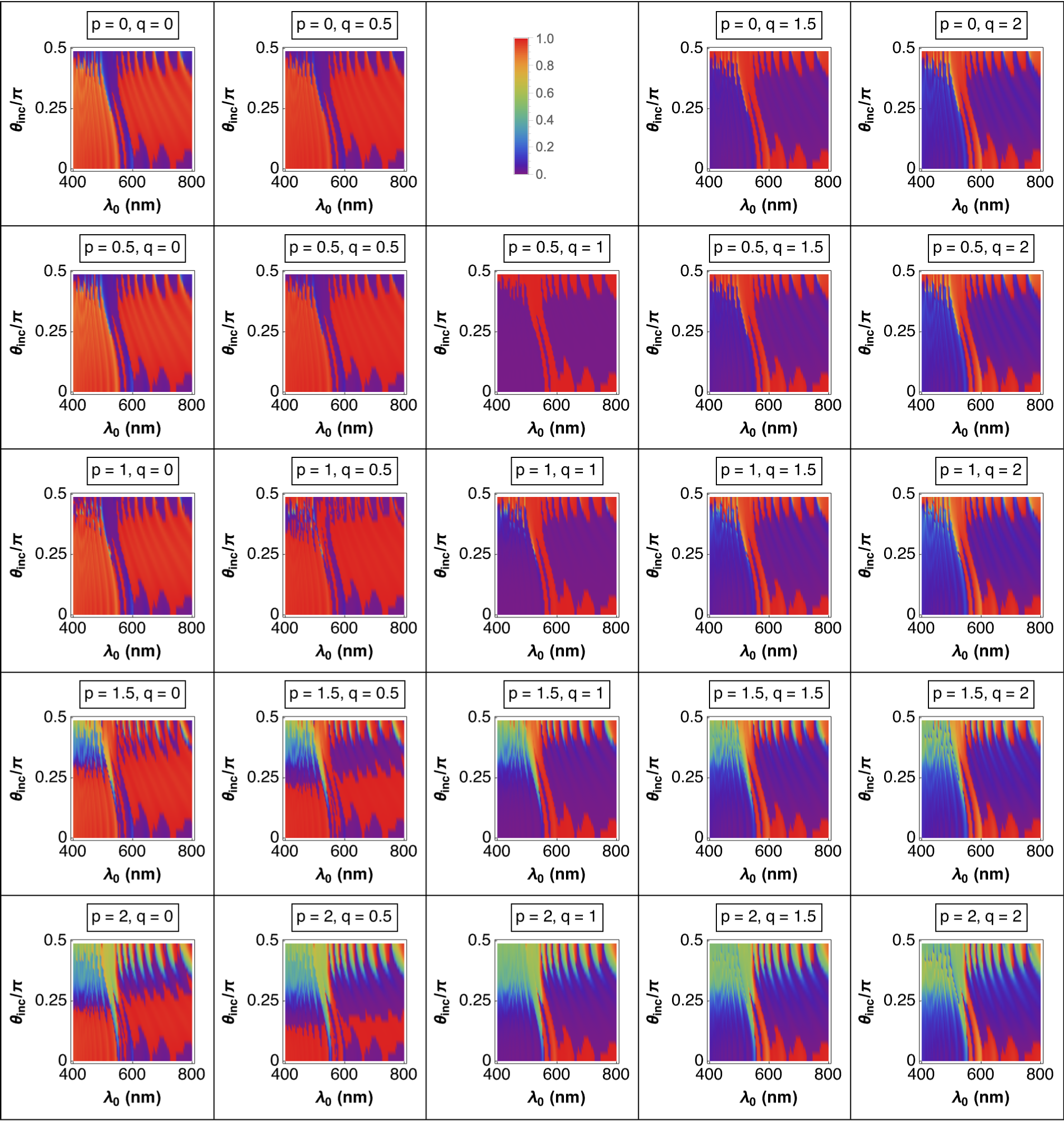} 
 \caption{\label{CSTF-twist-tra} 
 Same as Fig.~\ref{CSTF-tra} except that the  set of  five unit cells in the back of the CSTF are twisted by $\pi/2$ about the $+z$ axis.
}
\end{figure}

The central $90\deg$-twist defect clearly alters the DIGP landscape in Figs.~\ref{CSTF-twist-ref} and \ref{CSTF-twist-tra}
relative to Figs.~\ref{CSTF-ref} and \ref{CSTF-tra}.  Relative to the defect-free CSTF, the reflected and transmitted DIGP maps now show stronger distortion of the diagonal phase fronts, sharper local contrasts for some $(p,q)$ pairs, and a more pronounced rearrangement of the spectral-angular bands associated with the circular
Bragg regime. The systematic dependences on $p$ and $q$ remains evident, but the central $90\deg$-twist defect introduces additional phase mixing so that the contour patterns are no longer simple shifted replicas of their counterparts for the defect-free CSTF. In short, Figs.~\ref{CSTF-twist-ref} and \ref{CSTF-twist-tra}
 encode the same type of geometric-phase response as Figs.~\ref{CSTF-ref} and \ref{CSTF-tra}, but for a structurally chiral thin film with altered internal phase matching.
 In principle, the collection of DIGP maps in all four figures
could be used for inverse retrieval of   morphological and electromagnetic characteristics.

%%%%%%%%%%%%%%%%%% Figure 5 begins %%%%%%%%%%%%%%%%%%
%%%%%%%%%%%%%%%%%% Figure 5 ends %%%%%%%%%%%%%%%%%%
%%%%%%%%%%%%%%%%%% Figure 6 begins %%%%%%%%%%%%%%%%%%
%%%%%%%%%%%%%%%%%% Figure 6 ends %%%%%%%%%%%%%%%%%%

\subsubsection{Asymmetric chevronic sculptured thin film}\label{AChevSTF}
Different from the unit cell of a CSTF,
the unit cell of an asymmetric chevronic sculptured thin film (AChevSTF)
 comprises two columnar thin films (CTFs) of thicknesses $d_1$ and $d_2=D-d_1$ \cite{Monteiro,Amaya,Vepachedu,Rehman1,Rehman2}.
The  relative
permittivity dyadic of the first CTF is given by  
\begin{subequations}
\begin{equation}
\=\eps_1=\=S_{\rm y}(\chi_1)\.\left(\eps_{{\rm a}_1}\uz\uz+
\eps_{{\rm b}_1}\ux\ux+\eps_{{\rm c}_1}\uy\uy\right)
\.\=S_{\rm y}^{-1}(\chi_1)\,,
\label{eps1}
\end{equation}
where the principal relative permittivity scalars $\eps_{{\rm a}_1}$, $\eps_{{\rm b}_1}$, and $\eps_{{\rm c}_1}$ are
frequency dependent and the angle of inclination
$\chi_1\in(0,\pi/2]$. The relative
permittivity dyadic of the second CTF is similarly given by  
\begin{equation}
\=\eps_2=\=S_{\rm y}(\pi-\chi_2)\.\left(\eps_{{\rm a}_2}\uz\uz+
\eps_{{\rm b}_2}\ux\ux+\eps_{{\rm c}_2}\uy\uy\right)
\.\=S_{\rm y}^{-1}(\pi-\chi_2)\,,
\label{eps2}
\end{equation}
\end{subequations}
where $\eps_{{\rm a}_2}$, $\eps_{{\rm b}_2}$, and $\eps_{{\rm c}_2}$ are frequency dependent  and the   angle $\chi_2\in(0,\pi/2]$. 
The   AChevSTF comprises $N$ unit cells and is therefore of thickness $ND$ with structural period $D$. The absence of the dyadic $\=S_{\rm z}(\.)$
from Eqs.~\eqref{eps1} and \eqref{eps2} indicates that   AChevSTFs essentially have a two-dimensional (or nematic) morphology,
in contrast to the three-dimensional morphology of CSTFs. Also, whereas a CSTF is continuously non-homogeneous along the $z$ axis,
an AChevSTF is piecewise homogeneous along the same axis. The matrix $[\#M]$ for AChevSTFs is available elsewhere \cite{Vepachedu}.

\begin{figure}[!htb]
\centering
 \includegraphics[width=6.5cm]{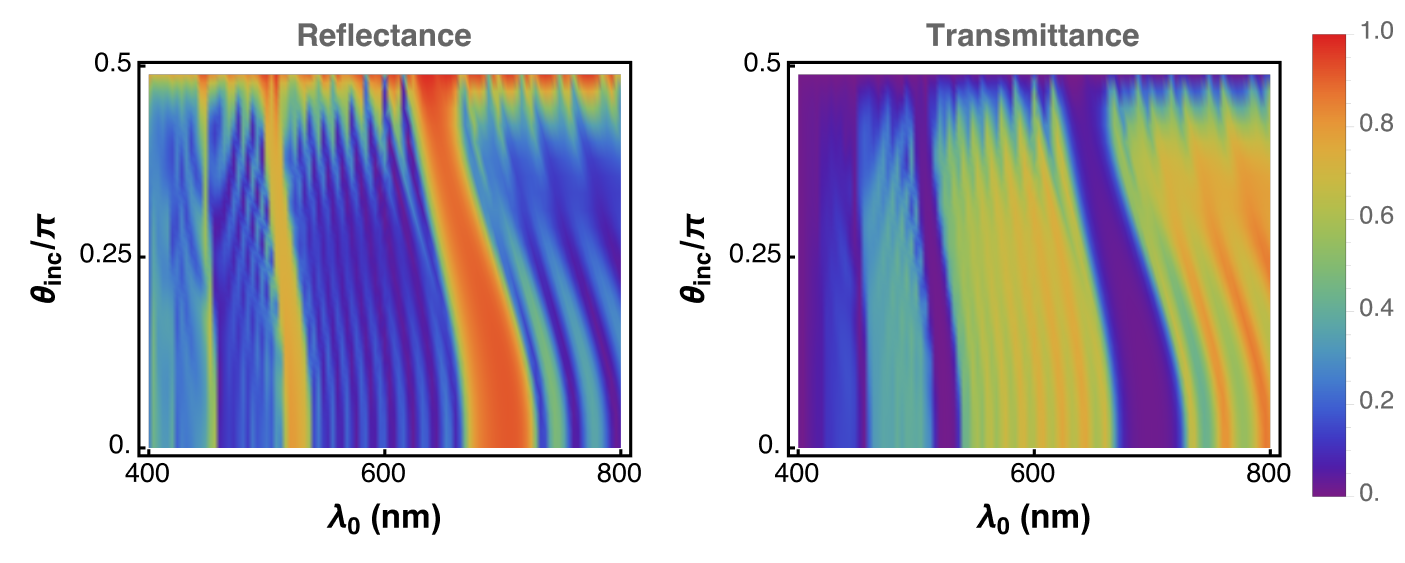} 
 \caption{\label{AChevSTF-remit} 
 Reflectance $R$ and transmittance $T$ as functions of
$\lambdao\in[400,800]$~nm and $\thetainc\in[0,\pi/2)$ when $\phiinc=\pi/4$, $\aas=1 + 0.3 i$, and $\aap=0.7 - 1.2 i$, for a 10-period-thick AChevSTF with constitutive parameters specified in Sec.~\ref{AChevSTF}.  
}
\end{figure}

Figure~\ref{AChevSTF-remit} provides  the reflectance $R$ and transmittance $T$   
as  functions of $\lambdao\in[400,800]$~nm and $\thetainc\in[0,\pi/2)$ when $\phiinc=45\deg$, $\aas=1 + 0.3 i$, and $\aap=0.7 - 1.2 i$,  calculated
for a 10-period-thick AChevSTF with the relative permittivity dyadic specified by Eqs.~\eqref{eps1} and \eqref{eps2}. The single-resonance Lorentzian
function applies with
\begin{itemize}
\item[(i)]
$p_{{\rm a}_1} = 1.024$, $p_{{\rm b}_1} =1.990$, $p_{{\rm c}_1} =1.458$, 
$\lambda_{{\rm a}_1}  = \lambda_{{\rm b}_1} =\lambda_{{\rm c}_1}  =360$~nm,   
$N_{{\rm a}_1}  = N_{{\rm b}_1}  =N_{{\rm c}_1} =500$, $\chi_1 = 46.36\deg$, and $d_1=81.25$~nm for the first CTF;
and
\item[(ii)]
$p_{{\rm a}_2} = 3.048$, $p_{{\rm b}_2} =3.287$, $p_{{\rm c}_2} =3.098$, $\lambda_{{\rm a}_2}  = \lambda_{{\rm b}_2} =\lambda_{{\rm c}_2}  =360$~nm,   
$N_{{\rm a}_2}  = N_{{\rm b}_2}  =N_{{\rm c}_2} =500$, $\chi_2 = 82.80\deg$, and $d_1=243.75$~nm for the second CTF
\end{itemize}
in the unit cell of the AChevSTF. Data were generated on
a grid of 101 regularly spaced values of $\lambdao$
and 45 regularly spaced values of $\thetainc$. All calculations were performed 
on Mathematica\texttrademark~version~14.0.0.0.

Two high-intensity ridges in the map of $R$ are mirrored by two low-intensity troughs in the map of $T$ in Fig.~\ref{AChevSTF-remit}. These features
are signatures of the regular, not circular, Bragg phenomenons of different orders \cite{Teitler1970,SSK1991,Slepyan1998,HW1999}. Both features
show a blueshift with increasing $\thetainc$ \cite{Vepachedu,Rehman1}, which is characteristic of periodic systems as also evident
 in Figs.~\ref{CSTF-remit} and \ref{CSTF-twist-remit}.
The center-wavelength and the bandwidth of the higher-order Bragg regime are, respectively, smaller and narrower
than those of the lower-order Bragg regime, as expected both theoretically and experimentally \cite{Vepachedu}. Fabry--P\'{e}rot resonances
are present between the two Bragg regimes as well as on the long-wavelength side of the lower-order Bragg regime, due to the finite
thickness of the AChevSTF.

%%%%%%%%%%%%%%%%%% Figure 7 begins %%%%%%%%%%%%%%%%%%
%%%%%%%%%%%%%%%%%% Figure 7 ends %%%%%%%%%%%%%%%%%%

\begin{figure}[!htb]
\centering
 \includegraphics[width=9.5cm]{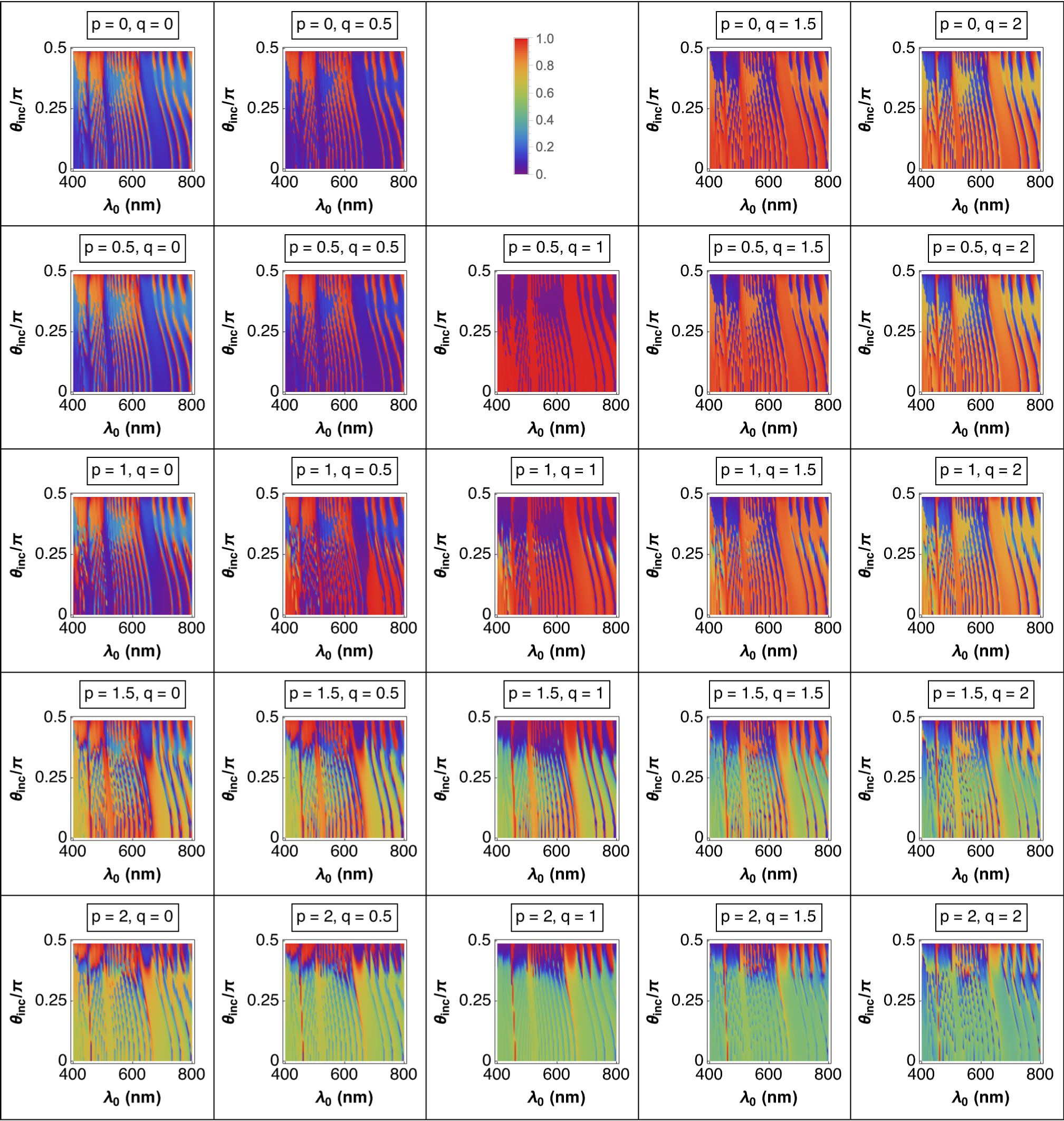} 
 \caption{\label{AChevSTF-ref} 
 $\Psipqref$ as a function of  $\lambdao\in[400,800]$~nm and $\thetainc\in[0,\pi/2)$ when $\phiinc=\pi/4$, $\aas=1 - 0.3 i$, and $\aap=0.7 + 1.2 i$, for a 10-period-thick AChevSTF with constitutive parameters specified in Sec.~\ref{AChevSTF}. Individual maps depict   $\Psipqref$  for $p\in\lec 0,0.5,1,1.5,2\ric$
 and $q\in\lec 0,0.5,1,1.5,2\ric$. Note that $\tPsi^{(0,1)}_{\rm ref}\equiv 0$.
}
\end{figure}

\begin{figure}[!htb]
\centering
 \includegraphics[width=9.5cm]{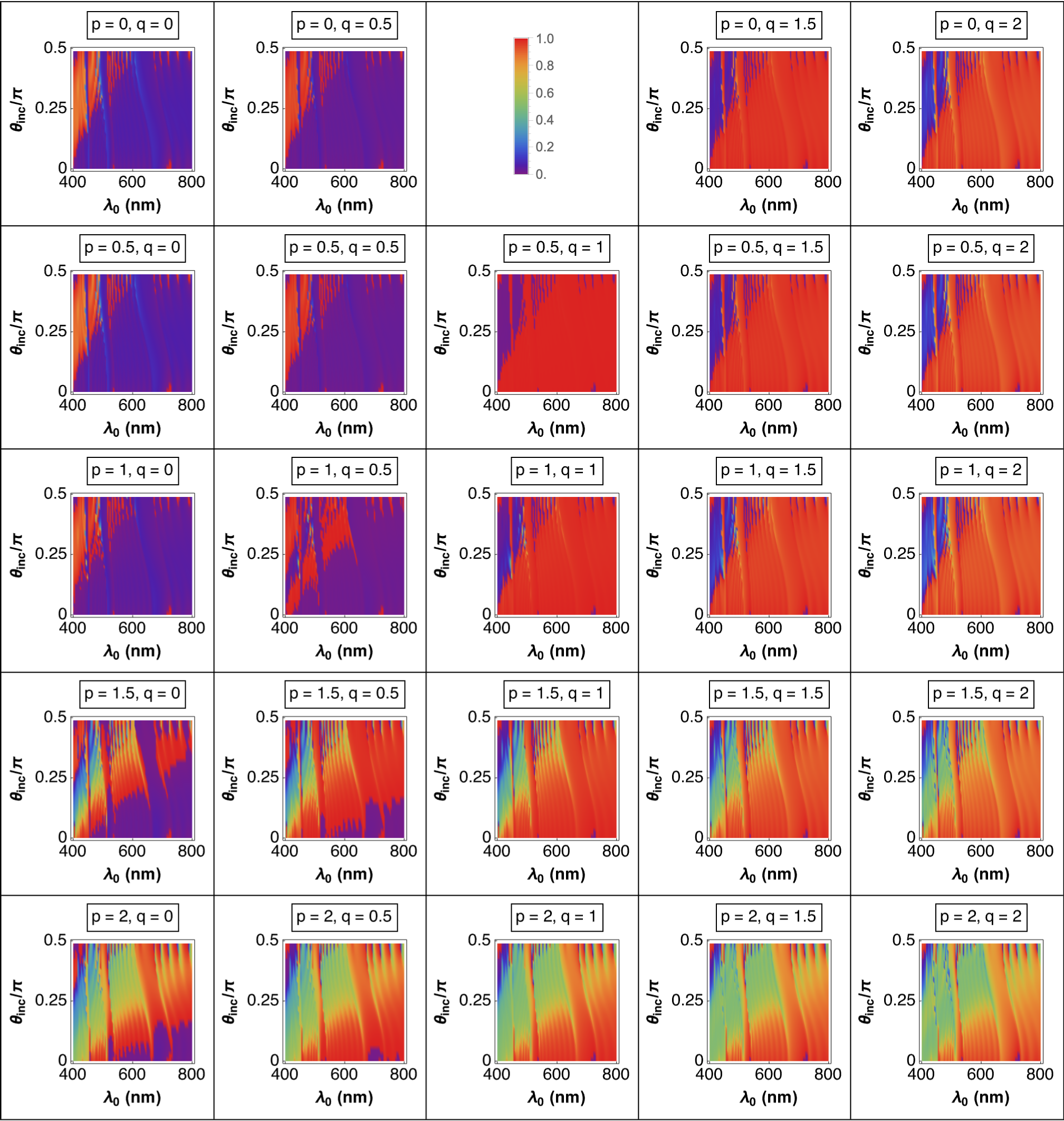} 
 \caption{\label{AChevSTF-tra} 
 $\Psipqtra$ as a function of  $\lambdao\in[400,800]$~nm and $\thetainc\in[0,\pi/2)$  when $\phiinc=\pi/4$, $\aas=1 - 0.3 i$, and $\aap=0.7 + 1.2 i$, for a 10-period
AChevSTF with constitutive parameters specified in Sec.~\ref{AChevSTF}. Individual maps depict   $\Psipqtra$  for $p\in\lec 0,0.5,1,1.5,2\ric$
 and $q\in\lec 0,0.5,1,1.5,2\ric$. Note that $\tPsi^{(0,1)}_{\rm tr}\equiv 0$.
}
\end{figure}

Figures~\ref{AChevSTF-ref} and \ref{AChevSTF-tra} provide maps of $\Psipqref$ and  $\Psipqtra$, respectively,  
as  in the $\lambdao$-$\theta$ plane spanning $\thetainc\in[0,90\deg)$ and $\lambdao\in[400,800]$~nm. These DIGP
maps show that the geometric phase is strongly modulated in the neighborhood of the Bragg ridges and troughs, 
while varying more smoothly away from the resonance regimes. The dependence on $p$ and $q$ remains systematic: 
varying $q$ produces an essentially linear phase offset, whereas varying $p$ reshapes the phase landscape by 
altering the effective latitude on the Poincar\'e sphere. Consequently,  maps for different $(p,q)$ pairs do not represent different 
physics in the usual sense, but rather different spinor gauges or projections of the same physical processes. Nevertheless, 
the family of DIGP maps provides complementary information about the  angles of inclination, thicknesses, and dielectric anisotropies
of the two CTFs in the unit cell of the AChevSTF. In principle, a collection of such maps could be used for inverse retrieval of 
morphological and electromagnetic parameters by combining forward transfer-matrix modeling with 
diverse parameter-estimation methods \cite{Tarantola,Kaipio,Hansen,Golub,Jolliffe}.

%%%%%%%%%%%%%%%%%% Figure 8 begins %%%%%%%%%%%%%%%%%%
%%%%%%%%%%%%%%%%%% Figure 8 ends %%%%%%%%%%%%%%%%%%
%%%%%%%%%%%%%%%%%% Figure 9 begins %%%%%%%%%%%%%%%%%%
%%%%%%%%%%%%%%%%%% Figure 9 ends %%%%%%%%%%%%%%%%%%

\subsubsection{Asymmetric chevronic sculptured thin film with central $90\deg$-twist defect}\label{AChevSTF-twist}

\begin{figure}[!htb]
\centering
 \includegraphics[width=6.5cm]{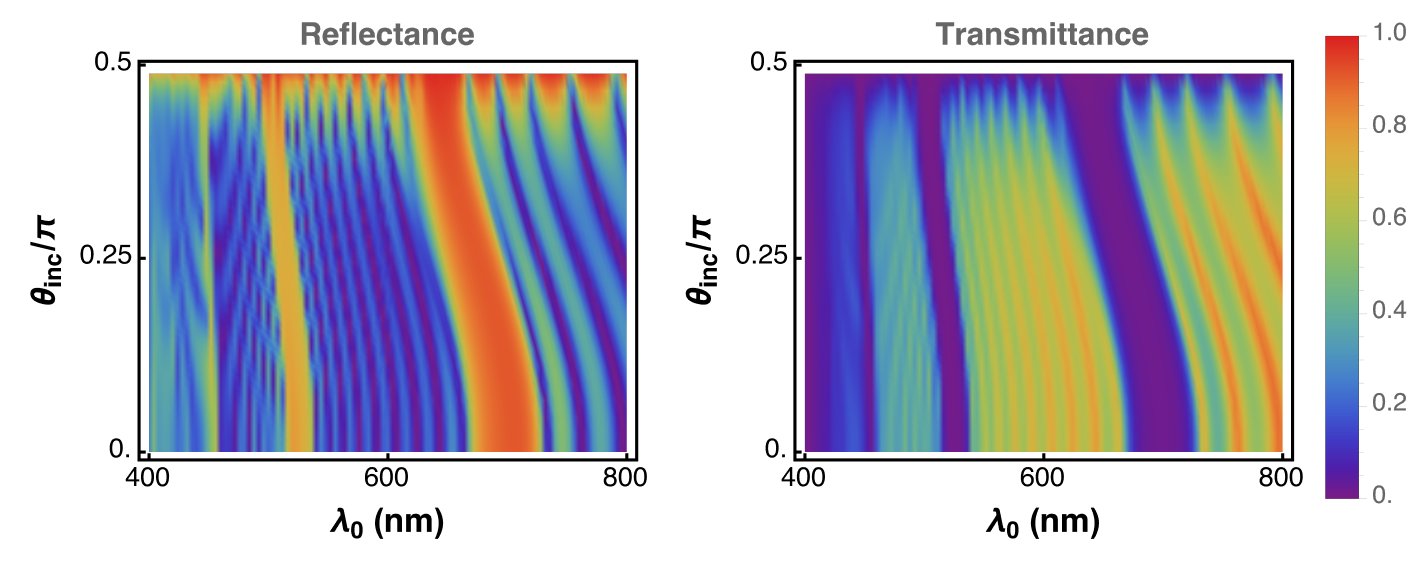} 
 \caption{\label{AChevSTF-twist-remit} 
 Same as Fig.~\ref{AChevSTF-remit} except that the set of five unit cells in the back of the AChevSTF are twisted by $\pi/2$ about the $+z$ axis.
}
\end{figure}

When one-half of an AChevSTF with an even number of unit cells is rotated by $90\deg$ about the $+z$ axis, the effect on $R$ and $T$ is almost unremarkable.
This is clear from a comparison of Fig.~\ref{AChevSTF-remit} for the defect-free AChevSTF with Fig.~\ref{AChevSTF-twist-remit} for the AChevSTF
with the central $90\deg$-twist defect. No bifurcation of either of the two Bragg regimes occurs, unlike for the CSTF in Sec.~\ref{CSTF-twist}. The most
prominent effect of the defect is a noticeable narrowing of both Bragg regimes for $\thetainc\lesssim45\deg$ and a noticeable
broadening of both Bragg regimes for $\thetainc\gtrsim45\deg$.

%%%%%%%%%%%%%%%%%% Figure 10 begins %%%%%%%%%%%%%%%%%%
%%%%%%%%%%%%%%%%%% Figure 10 ends %%%%%%%%%%%%%%%%%%

%%%%%%%%%%%%%%%%%% Figure 11 begins %%%%%%%%%%%%%%%%%%
 \begin{figure}[!htb]
\centering
 \includegraphics[width=9.5cm]{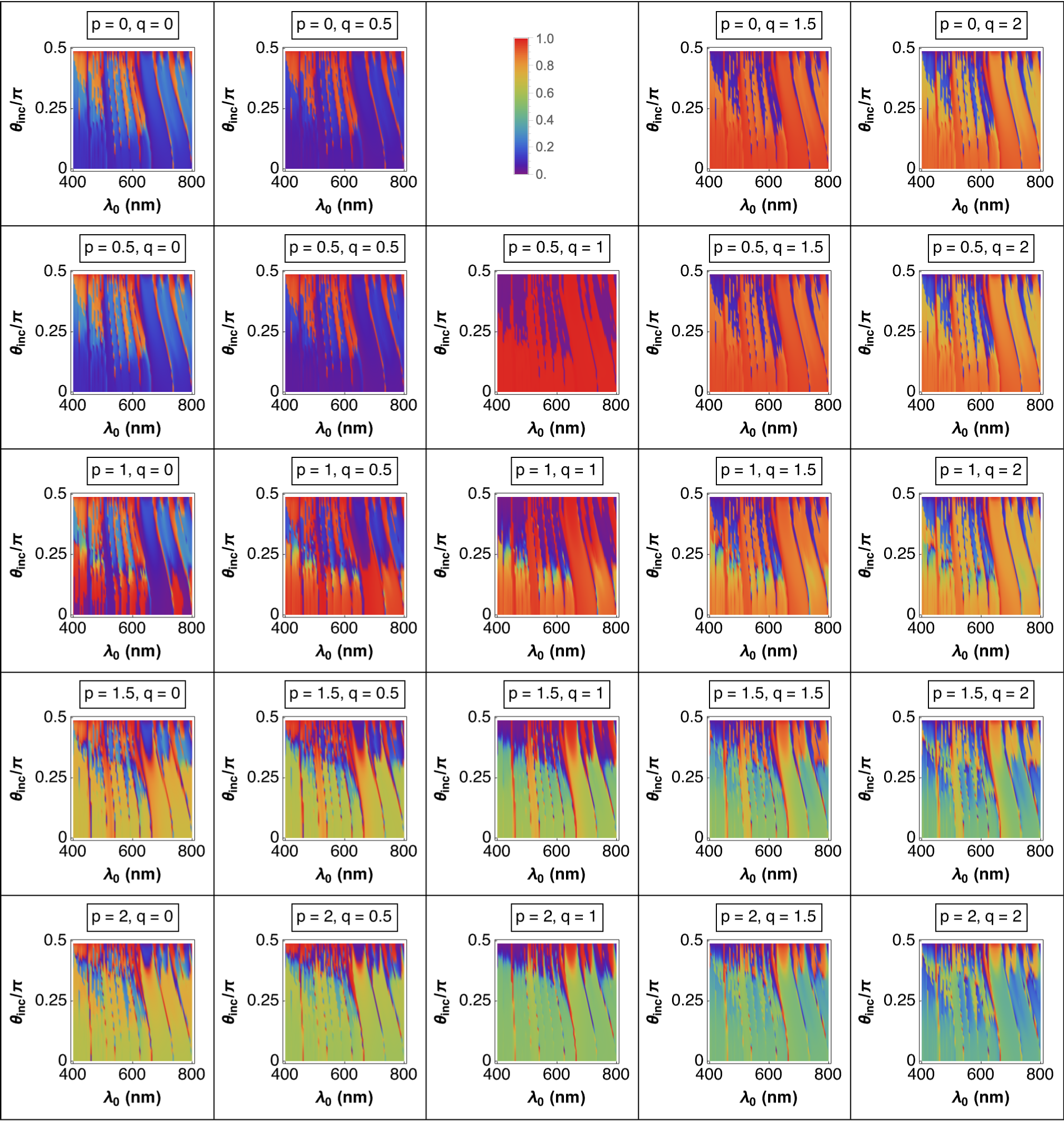} 
 \caption{\label{AChevSTF-twist-ref} 
Same as Fig.~\ref{AChevSTF-ref} except that the set of five unit cells in the back of the AChevSTF are twisted by $\pi/2$ about the $+z$ axis.
}
\end{figure}
%%%%%%%%%%%%%%%%%% Figure 11 ends %%%%%%%%%%%%%%%%%%
%%%%%%%%%%%%%%%%%% Figure 12 begins %%%%%%%%%%%%%%%%%%
 \begin{figure}[!htb]
\centering
 \includegraphics[width=9.5cm]{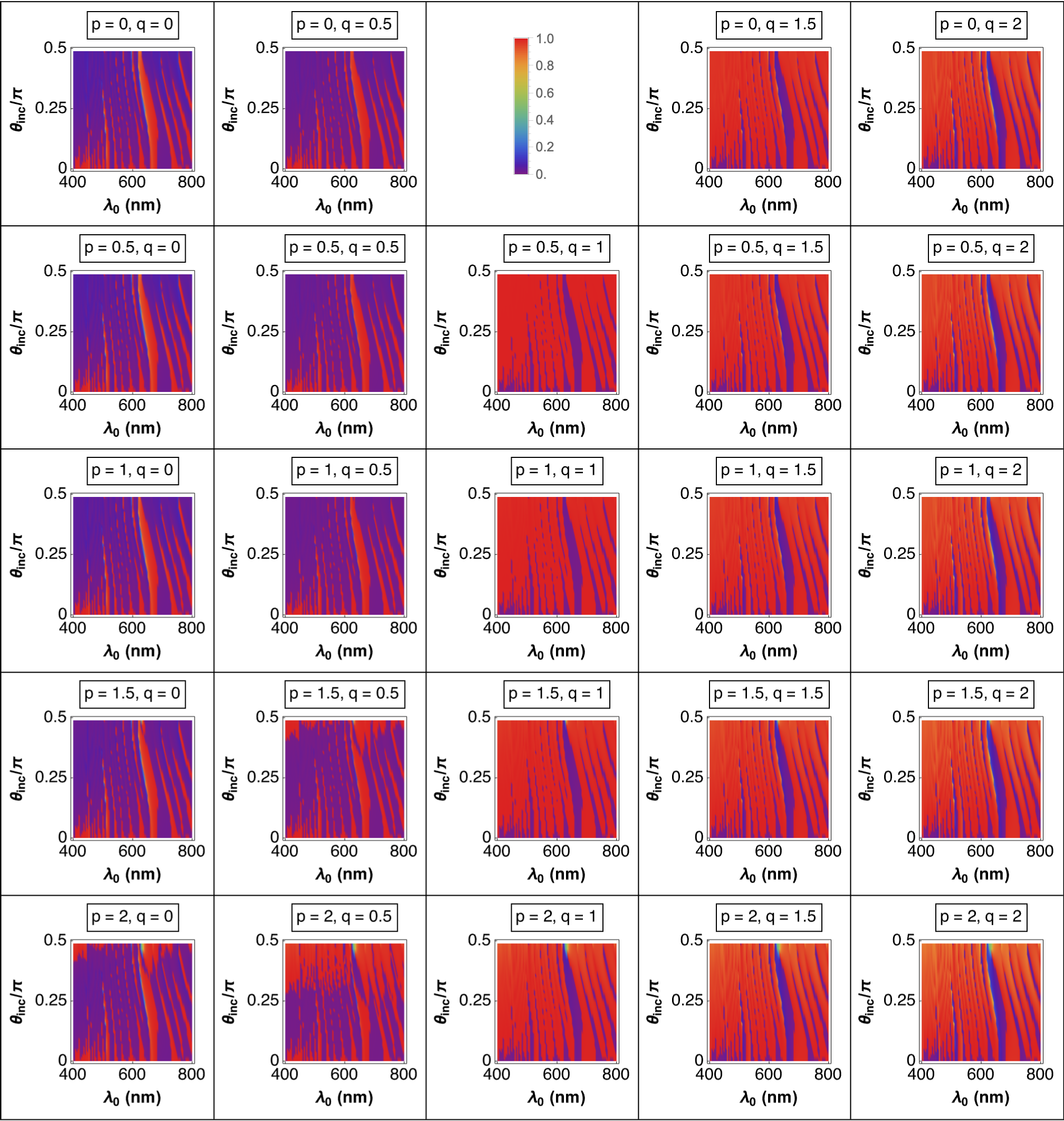} 
 \caption{\label{AChevSTF-twist-tra} 
 Same as Fig.~\ref{AChevSTF-tra} except that the  set of  five unit cells in the back of the AChevSTF are twisted by $\pi/2$ about the $+z$ axis.
}
\end{figure}
%%%%%%%%%%%%%%%%%% Figure 12 ends %%%%%%%%%%%%%%%%%%

But the maps of $\Psipqref$ and  $\Psipqtra$ in Figs.~\ref{AChevSTF-twist-ref} and \ref{AChevSTF-twist-tra}, respectively, for the AChevSTF with
the central $90\deg$-twist defect look very different from those in Figs.~\ref{AChevSTF-ref} and \ref{AChevSTF-tra} for the defect-free AChevSTF.
Sharp features are present  for $\thetainc \lesssim 45\deg$ in all four sets of DIGP maps. The difference is that the density of those sharp features has either
increased or decreased with the introduction of the central $90\deg$-twist defect, depending on values of $p$ and $q$. Thus, the DIGP maps
provide significant indications of the presence of   the central $90\deg$-twist defect, whereas the changes in the reflectance and transmittance maps
are unremarkable. 

A straightforward conclusion at this stage of research is that DIGP maps can  deliver more and different information about the structure and
electromagnetic constitution of a thin film than reflectance maps and transmittance maps.

\subsection{Plane-wave scattering by three-dimensional objects}\label{sec:3D}

The concept of geometric phase in electromagnetics arose to quantitate the interaction between two uniform plane waves propagating in the same
direction with the same frequency \cite{Pancha1}, as discussed via Eq.~\eqref{two} in Sec.~\ref{novelty}. From then onwards, it was  applied to
study plane-wave  transmission characteristics of planar structures such as thin films \cite{Bhandari,JishaReview,Rao2026} and effectively planar 
structures such as metasurfaces \cite{FarazReview,ChenReview,YuReview}. Such application is natural, when comparing a transmitted plane wave
with the incident plane wave that engendered it because the geometric phase encodes the action of the planar structure on the incident plane wave.
Plus, the transmitted plane wave moves in the same direction as the incident plane wave, by virtue of    Ibn Sahl's law of refraction (commonly attributed to 
Willebrord   Snel van~Royen) \cite{Rashed}.

A maverick looking at Pancharatnam's paper could easily conclude that the two plane waves labeled $1$ and $2$ in Eq.~\eqref{two} need not be co-propagating.
Only the locations $(\alpha_1,\beta_1)$ and  $(\alpha_2,\beta_2)$ on the Poincar\'e sphere  of two plane waves are needed to create the
Pancharatnam phase $\Psis_{12}$. Therefore,
the concept of geometric phase can be extended from its traditional use in comparing plane waves and polarization states to even plane-wave scattering   by a 
three-dimensional object of finite volume \cite{LakhtakiavdH}.

This extension is justified by a structural analogy as follows.  Two plane waves are not identical when their locations on the
Poincar\'e sphere differ, and the relative orientation of their polarization spinors defines a geometric phase. A three-dimensional object does not produce 
a scattered plane wave. Instead, it produces a continuum of outgoing waves, each associated with a direction. In the far zone, every outgoing component is locally transverse and can be treated as a plane wave propagating in its own direction. Its polarization state can be located on the Poincar\'e sphere. This makes it meaningful to compare  
 the far-zone scattered  wave in a certain direction  with
the  incident plane wave   through the geometric-phase concept. The geometric-phase portrayal of plane-wave
scattering created thereby does not replace the conventional scattering pattern \cite{BSU,BH83,VLV,Elsherbeni} but can  provide additional information on the scattering object 
\cite{LakhtakiavdH}. The DIGP concept is a further development.

\subsubsection{{Boundary-value problem}}\label{sec:bvp}

Let all space $\cal V$ be divided into two   disjoint  regions $\Vint$ and $\Vext$.
The three-dimensional object occupies the region $\Vint$, which is bounded in all directions by the closed surface $\cal S$ 
and filled with an electromagnetic medium different from free space. The origin of the coordinate system lies inside $\Vint$. Extending
to infinity in all directions, the
region $\Vext$ is vacuous. 

Without any loss of generality, the incident plane wave can be taken to propagate along the $+z$ axis; hence, Eqs.~\eqref{eqEi-lin} and \eqref{eqEi-circ}
with $\thetainc=0$ and $\phiinc=3\pi/2$ can be used \cite{LakhtakiavdH}. Outside the smallest sphere with its center at the origin and
circumscribing $\Vint$, the scattered electric field phasor can be written as
 \begin{eqnarray}
 \nonumber
 \#E\sca(\#r)&=& \sum_{s\in\left\{e,o\right\}}\sum^{\infty}_{n=1}\sum^n_{m=0}\Dmn\les
 A\smn\three\,{{\#M}\smn\three}(\ko\#r)
 \right.
 \\[8pt]
&&
 \left.
 +  B\smn\three\,{{\#N}\smn\three}(\ko\#r)\ris\,,\quad r>a\,.
 \label{Esca}
 \end{eqnarray}
Here, $a$ is the radius of the circumscribing sphere; the coefficient
\begin{equation}
\displaystyle{
\Dmn=(2-\delta_{\rm m0}){(2n+1)(n-m)!\over 4n(n+1)(n+m)!}
}
\end{equation}
employs the Kronecker delta $\delta_{\rm mm^\prime}$; the vector spherical wavefunctions
${{\#M}\smn^{(3)}}(\ko\#r)$ and ${{\#N}\smn^{(3)}}(\ko\#r)$   \cite{Stratton1941,MF1953} are provided in Appendix 1;
and $A\smn\three$ and $B\smn\three$ are unknown coefficients of expansion.

The determination of $\#E\sca(\#r)$---and, therefore, of $A\smn\three$ and $B\smn\three$---for specified $\Ei$ requires the  solution of a 
boundary-value problem. The interested reader is referred to
a host of analytical
and numerical techniques \cite{BSU,Lakh-MoM,Yurkin,Chew,SIE,Sertel,Lakh-EBCM,Vavilin,Elsherbeni,FEM-BEM,MAS}  for that purpose. 

\subsubsection{Direction-dependent DIGP of the far-zone scattered field}
 In the far zone (i.e., as $\ko r\to \infty$), the scattered electric field  phasor may be approximated as \cite{Saxon,Kristensson}
\begin{equation}
\#E\sca(r\hat{\#r})\approx  \les\Thetasca(\hat{\#r}) \utheta
 +\Phisca(\hat{\#r}) \uphi\ris\frac{\exp(i{\ko}r)}{\ko r}\,,
\label{def-Esca-far}
\end{equation}
where
\begin{subequations}
\begin{eqnarray}
\nonumber
\Thetasca(\hat{\#r}) 
&=&\sum_{s\in\left\{e,o\right\}}\sum^{\infty}_{n=1}\sum^n_{m=0}\lec
(-i)^n\,\Dmn \les-iA\smn\three\,{f\smn}(\theta,\phi)
\right.\right.
\\[8pt]
&&\quad\left.\left.+B\smn\three\,{g\smn}(\theta,\phi)\ris \ric
\label{def-Thetasca}
\end{eqnarray}
and
\begin{eqnarray}
\nonumber
 \Phisca(\hat{\#r}) 
&=&\sum_{s\in\left\{e,o\right\}}\sum^{\infty}_{n=1}\sum^n_{m=0}\lec
(-i)^n\,\Dmn \les iA\smn\three\,{g\smn}(\theta,\phi)
\right.\right.
\\[8pt]
&&\quad\left.\left.+B\smn\three\,{f\smn}(\theta,\phi)\ris \ric
\label{def-Phisca}
\end{eqnarray}
\end{subequations}
involve the functions $f\smn(\theta,\phi)$ and $g\smn(\theta,\phi)$
 defined in   Appendix~1. 
 
Knowing $\Thetasca(\hat{\#r}) $ and $ \Phisca(\hat{\#r})$, we can determine
the direction-dependent Stokes parameters of the scattered wave as \cite{LakhtakiavdH}
\begin{equation}
\left.\begin{array}{l}
\szero^{\rm sca}(\ur)
=  \vert {\Thetasca(\ur)}\vert^2+ \vert {\Phisca(\ur)}\vert^2
\\[5pt]
s_{1}^{\rm sca}(\ur) =  \vert {\Thetasca(\ur)}\vert^2- \vert {\Phisca(\ur)}\vert^2
\\[5pt]
s_{2}^{\rm sca}(\ur) 
=-2\,{\rm Re}\les   {\Phisca(\ur) \,\Thetasca^\ast(\ur)}\ris
\\[5pt]
s_{3}^{\rm sca}(\ur) 
=-2\,{\rm Im}\les   {\Phisca(\ur)\, \Thetasca^\ast(\ur)}\ris
\end{array}\right\}\,.
\end{equation}
These quantities can be used in Eq.~\eqref{def-alphabeta}
to obtain the direction-dependent  longitude $\alpha\sca(\ur)$ and latitude $\beta\sca(\ur)$,
whence the direction-dependent DIPS $\bpsipqsca$
may be obtained using Eq.~\eqref{def-Ppq-0}.

The direction-dependent DIGP of the scattered wave with respect to the incident plane wave
may be then computed as
\begin{equation}
\label{def-DIGP-sca}
\Psipqsca=\Arg\lec{\bpsipqinc}^\dag\.{\bpsipqsca}\ric  \,,
\end{equation}
following Eq.~\eqref{def-DIGP}.

\subsubsection{Homogeneous isotropic chiral sphere}
For the sake of illustration, let $\Vint$ be a sphere of radius $a$ occupied by homogeneous isotropic chiral material \cite{Bohren1974,Hector2025}
with frequency-domain constitutive relations
\begin{equation}
\left.
\begin{array}{l}
    \#{D}(\#r) = \epso\, \epsr  \, \#{E}(\#r) + i \, \varkappa \sqrt{\epso\muo} \, \#{H} (\#r)
    \\[5pt]
    \#{B} (\#r)= \muo \, \mur \,\#{H}(\#r) - i \, \varkappa  \sqrt{\epso \muo} \, \#{E}(\#r)
\end{array}
\right\}\,,\quad \#r\in\Vint\,,
\label{def-chiral}
\end{equation}
where $\epsr \in\mathbb{C}$ is the frequency-dependent relative permittivity scalar,  $\mur \in\mathbb{C}$ is the frequency-dependent  relative 
permeability scalar, and  $\varkappa \in\mathbb{C}$ is the frequency-dependent  Tellegen chirality parameter. Data were generated  on
a grid of 37 regularly spaced values of $\theta\in[0,\pi]$
and 37 regularly spaced values of $\phi\in[0,2\pi)$ 
 for   $\ko a =5$,  $\epsr=3(1+0.1i)$, $\mur=1.1 (1+0.05i)$, $\varkappa =0.5(1+0.2i)$, when
$\aas=0.5-0.18 i$ and $\aap=0.6 + 1.2 i$. The frequency was kept fixed.

\begin{figure}[!htb]
\centering
 \includegraphics[width=4.5cm]{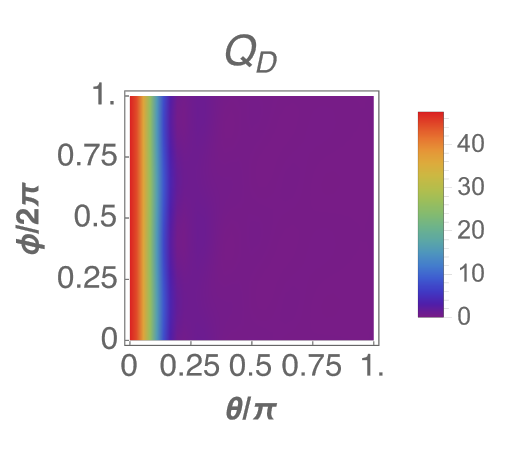} 
 \caption{\label{sphere-QD} 
 $\QD(\ur)$ as a function of   $\theta\in[0,\pi)$ and $\phi\in[0,2\pi)$ 
 calculated for a homogeneous isotropic chiral sphere of normalized
 size $\ko a =5$, when $\epsr=3(1+0.1i)$, $\mur=1.1 (1+0.05i)$, $\varkappa =0.5(1+0.2i)$,
   $\aas=0.5-0.18 i$, and $\aap=0.6 + 1.2 i$.}
\end{figure}

The chief standard measure of plane-wave scattering by a three-dimensional object is
the differential scattering efficiency \cite{Saxon,RCS,LakhtakiavdH}
\begin{equation}
\label{sigmad-def}
\QD(\ur)=\frac{4}{\ko^2a^2} \dfrac{\vert\Thetasca(\ur)\vert^2+\vert\Phisca(\ur)\vert^2}
{\vert\aas\vert^2+\vert\aap\vert^2} \,
\end{equation}
along any radial direction. Figure~\ref{sphere-QD} shows $\QD$ as a function of   $\theta$ and $\phi$. 
Whereas $\QD$ is dominant in the forward sector $\theta\lesssim30\deg$,
it is uniformly small for $\theta\in[35\deg,180\deg]$; furthermore, the variation of $\QD$ with $\phi$ is very weak.

%%%%%%%%%%%%%%%%%% Figure 13 begins %%%%%%%%%%%%%%%%%%
%%%%%%%%%%%%%%%%%% Figure 13 ends %%%%%%%%%%%%%%%%%%

%%%%%%%%%%%%%%%%%% Figure 14 begins %%%%%%%%%%%%%%%%%%
 \begin{figure}[!htb]
\centering
 \includegraphics[width=9.5cm]{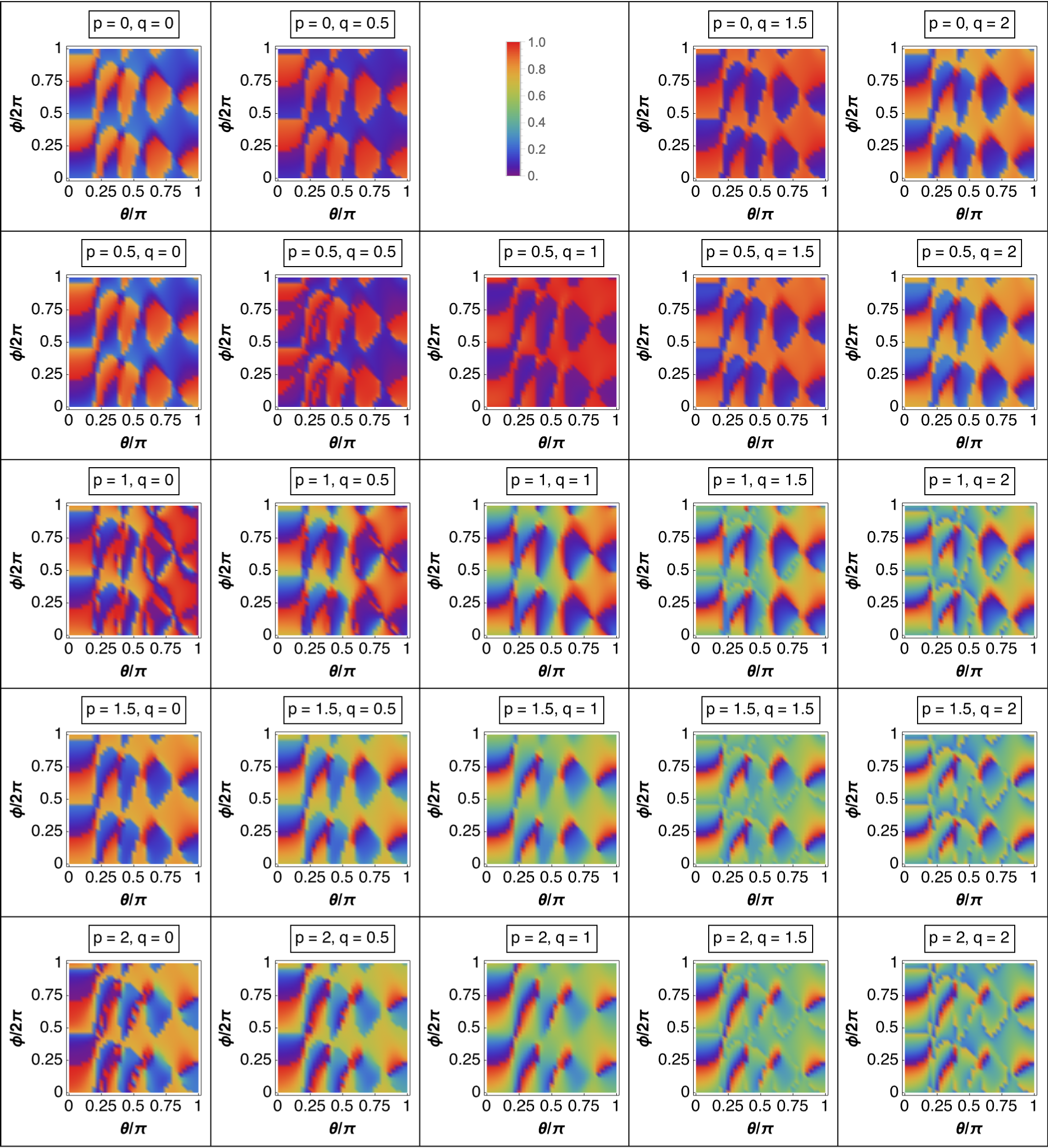} 
 \caption{\label{sphere-DIGP} 
 $\Psipqsca$ as a function of   $\theta\in[0,\pi)$ and $\phi\in[0,2\pi)$ for a homogeneous isotropic chiral sphere of normalized
 size $\ko a =5$, when $\epsr=3(1+0.1i)$, $\mur=1.1 (1+0.05i)$, $\varkappa =0.5(1+0.2i)$,
   $\aas=0.5-0.18 i$, and $\aap=0.6 + 1.2 i$.
 Individual maps depict   $\Psipqsca$  for $p\in\lec 0,0.5,1,1.5,2\ric$
 and $q\in\lec 0,0.5,1,1.5,2\ric$. Note that $\tPsi^{(0,1)}_{\rm sca}\equiv 0$.
}
\end{figure}
%%%%%%%%%%%%%%%%%% Figure 14 ends %%%%%%%%%%%%%%%%%%

In contrast, $\Psipqsca$ has strong dependencies on both $\theta$ and $\phi$, as
becomes clear from the DIGP maps in Fig.~\ref{sphere-DIGP}.  
Except for $\tPsi^{(0,1)}_{\rm sca}(\ur)\equiv 0$, the DIGP maps
 contain multiple bands, lobes, extremums, and phase reversals, indicating that the DIGP maps capture phase-dependent information that is largely absent from 
 the $\QD$ map. Systematic changes occur as the indexes $p$ and $q$  are varied. Increasing $q$ primarily shifts and reorganizes the phase features, 
 consistent with its longitudinal action in the Poincar\'e sphere.  Variation of $p$, however, modifies the effective latitude on the Poincar\'e sphere and therefore reshapes the overall angular landscape, changing the positions, orientations, and contrasts of diverse features. The most important prospective application of DIGP maps is 
 therefore inverse scattering \cite{LakhtakiavdH}. If systematic trends in DIGP maps  with shape, constitutive parameters, and boundary conditions
are identified, they may be used to classify and identify diverse scatterers, be they even non-spherical and non-homogeneous.

\subsection{Individual DIGP maps or collections of DIGP maps?}\label{PCA}
Should an individual DIGP map for a specific $(p,q)$ pair suffice for inverse-scattering problems or will a collection of DIGP maps for diverse $(p,q)$ pairs,
all calculated from the same set of polarimetric measurements, be needed? Although the real answer(s) to that question will require further research,
a preliminary indication was provided by a principal component analysis (PCA)  of the DIGP maps in Figs.~\ref{CSTF-ref},
\ref{CSTF-tra}, \ref{CSTF-twist-ref}, \ref{CSTF-twist-tra}, \ref{AChevSTF-ref},
\ref{AChevSTF-tra}, \ref{AChevSTF-twist-ref}, \ref{AChevSTF-twist-tra}, and \ref{sphere-DIGP}.

PCA is a statistical method that transforms a set of correlated variables into a smaller number of uncorrelated variables 
called principal components \cite{Jolliffe}. Each of the 24 DIGP
maps in each of the nine figures was treated as one observation, with the DIGP at each  grid point 
serving as the variable. Whereas the grid for Figs.~\ref{CSTF-ref},
\ref{CSTF-tra}, \ref{CSTF-twist-ref}, \ref{CSTF-twist-tra}, \ref{AChevSTF-ref},
\ref{AChevSTF-tra}, \ref{AChevSTF-twist-ref}, and \ref{AChevSTF-twist-tra} comprised $101\times45$ points,
the grid for Fig.~\ref{sphere-DIGP} comprised $37\times37$ points.
For each figure, the 24 DIGP maps were mean-centered and decomposed into orthogonal principal
components. All calculations were performed using the functions \textsf{PrincipalComponents} and \textsf{Variance}
on Mathematica\texttrademark~version~14.0.0.0.

PCA identifies the directions in this high-dimensional pixel space along which the maps exhibit the greatest variance. The first principal component (PC1)
explains the largest fraction of the total variance, the second (PC2) explains the largest remaining fraction, and so on. The corresponding component scores indicate how strongly each DIGP map expresses those dominant patterns. Thus, PCA provides an objective way to determine whether the changes across the DIGP maps in a collection 
are mainly associated with $p$, with $q$, or with their interaction. It also reveals whether many DIGP maps are effectively redundant and can help reduce the dimensionality of the data for classification or inverse-scattering analyses.

Table~\ref{tab:pca_variance} demonstrates that the DIGP maps within each figure are well represented by a low-dimensional principal-component structure. PC1 is consistently dominant, accounting for 60.7\% to 94.1\% of the variance across the nine figures. Its mean contribution is 75.1\%, indicating that a single common pattern explains most of the map-to-map variation. PC1 is especially pronounced for Fig.~\ref{AChevSTF-twist-tra} (94.1\%), whereas Fig.~\ref{CSTF-twist-tra} exhibits the lowest PC1 contribution (60.7\%), suggesting comparatively greater structural diversity in the latter set.

The second component contributes a further 3.9\% to 17.0\%, with the largest contribution occurring in Fig.~\ref{CSTF-twist-tra}. Consequently, the cumulative variance explained by PC1 and PC2 ranges from 77.7\% to 98.0\%, averaging approximately 85\%. This result confirms that two components capture the major variation in every figure. The third component is smaller, ranging from 1.5\% to 8.1\%, but raises the cumulative explained variance to between 85.8\% and
99.5\%. Figures~\ref{CSTF-tra} and \ref{AChevSTF-twist-tra} show particularly compact representations, with the first three components explaining 
95.7\% and 99.5\% of variance, respectively. Conversely, Figs.~\ref{CSTF-twist-ref} and \ref{CSTF-twist-tra} retain the greatest residual variation after three components. Overall, these results establish substantial redundancy among the 24 DIGP maps per figure and support the use of the first two or three principal components for reduced-dimensionality analysis, visualization, and interpretation. The consistently high cumulative values also indicate that the dominant variability is organized rather than dispersed across many independent map features.

%%%%%%%%%%%%%%%%%%%%% Table 2 begins %%%%%%%%%%%%%%%%%%%%%%
\begin{table}[htbp]
\centering
\caption{Variance explained by the first three principal components for the DIGP maps in each of nine figures.}
\label{tab:pca_variance}
\begin{tabular}{c|@{\hskip 4pt}c@{\hskip 10pt}c@{\hskip 10pt}c|@{\hskip 10pt}c@{\hskip 10pt}c}
\hline
Figure & PC1 (\%) & PC2 (\%) & PC3 (\%) & PC1$+$PC2 (\%) & PC1$+$PC2$+$PC3 (\%) \\
\hline
\ref{CSTF-ref}  & 64.9 & 13.7 & 8.4 & 78.6 & 87.0 \\
\ref{CSTF-tra}  & 82.0 & 9.7 &  4.0 & 91.7 & 95.7\\
\ref{CSTF-twist-ref}  & 63.9 & 14.0 & 8.6 & 77.9 & 86.5 \\
\ref{CSTF-twist-tra}  & 60.7 & 17.0 &  8.1 & 77.7 & 85.8 \\
\ref{AChevSTF-ref}  & 68.9 & 12.8 & 6.2 & 81.7 & 87.9 \\
\ref{AChevSTF-tra}  & 81.1 & 8.9 &  5.3 & 90.0 & 95.3 \\
\ref{AChevSTF-twist-ref} & 64.9 & 16.1 & 6.0 & 81.0 & 87.0 \\
\ref{AChevSTF-twist-tra}& 94.1 & 3.9 &  1.5 & 98.0 & 99.5 \\
\ref{sphere-DIGP} & 78.3 & 7.3 &  6.0 & 85.6 & 91.6 \\
\hline
\end{tabular}
\end{table}
%%%%%%%%%%%%%%%%%%%%% Table 2 ends %%%%%%%%%%%%%%%%%%%%%%

The component correlations in Table~\ref{tab:pca_correlations} identify the roles of the indexes $p$ and $q$. 
For every figure, PC1 has its largest correlation with $q$, with the magnitude of the Pearson correlation
coefficient \cite{Pearson} ranging from $0.34$ in Fig.~\ref{sphere-DIGP}
 to $0.87$ in Fig.~\ref{AChevSTF-twist-tra}. The   correlation of PC1 with $p$ is substantially weaker, ranging from 0.09 to 0.51, 
 indicating that the dominant mode of variation among the DIGP maps is primarily governed by $q$. This pattern is especially clear in 
six of the nine figures,   with the relevant Pearson correlation
coefficient    exceeding $0.69$.

%%%%%%%%%%%%%%%%%%%%% Table 3 begins %%%%%%%%%%%%%%%%%%%%%%
\begin{table}[htbp]
\centering
\caption{Magnitudes of Pearson correlation coefficients of PCA scores with indexes $p$ and $q$.}
\label{tab:pca_correlations}
\begin{tabular}{c|@{\hskip 4pt}c@{\hskip 10pt}c|@{\hskip 10pt}c@{\hskip 10pt}c|@{\hskip 10pt}c|@{\hskip 10pt}c}
\hline
Figure & $\mathrm{PC1}$\&$p$ & $\mathrm{PC1}$\&$q$ & $\mathrm{PC2}$\&$p$ & $\mathrm{PC2}$\&$q$ & Stronger   & Strongest   \\
  &  &   &   &   &   association &   association \\
\hline
\ref{CSTF-ref}   & 0.09 & 0.75 & 0.04 & 0.35 & $q$ with PC1; $q$ with PC2 & $q$ with PC1\\
\ref{CSTF-tra}   & 0.18 & 0.84 & $0.33^\star$ & $0.24^\star$ & $q$ with PC1; $p$ with PC2 & $q$ with PC1\\
\ref{CSTF-twist-ref}   & 0.11 & 0.77 & 0.03    & 0.31 & $q$ with PC1; $q$ with PC2& $q$ with PC1 \\
\ref{CSTF-twist-tra}  & 0.18 & 0.84 &$0.32^\star$ & $0.24^\star$ & $q$ with PC1; $p$ with PC2 & $q$ with PC1 \\
\ref{AChevSTF-ref}  & 0.10 & 0.69 & 0.02 & 0.47 & $q$ with PC1; $q$ with PC2& $q$ with PC1 \\
\ref{AChevSTF-tra}  & 0.51 & 0.71 & 0.74 & 0.31 & $q$ with PC1; $p$ with PC2 & $p$ with PC2 \\
\ref{AChevSTF-twist-ref} & 0.09 & 0.69 & 0.07    & 0.48 & $q$ with PC1; $q$ with PC2 & $q$ with PC1 \\
\ref{AChevSTF-twist-tra}  & 0.09 & 0.87 & $0.39^\star$ & $0.04^\star$ & $q$ with PC1; $p$ with PC2 & $q$ with PC1 \\
\ref{sphere-DIGP} & 0.25 & 0.34 & $0.18^\star$    & $0.43^\star$ & $q$ with PC1; $q$ with PC2 & $q$ with PC2 \\
\hline
\end{tabular}
\vspace{2mm}
\begin{minipage}{0.93\linewidth}
\small
\end{minipage}
\end{table}
%%%%%%%%%%%%%%%%%%%%% Table 3 ends %%%%%%%%%%%%%%%%%%%%%%

The association structure of PC2 is more variable.  
PC2 is more strongly related to $p$ in four figures. In contrast, the remaining five figures show stronger absolute PC2 correlations with $q$. 
%The negative values of the Pearson correlation coefficient between PC2
%and $q$ for in Figs.~\ref{CSTF-tra}, \ref{CSTF-twist-tra}, \ref{AChevSTF-twist-tra}, 
%and \ref{sphere-DIGP} indicate an inverse relationship, rather than an absence of 
%$q$-dependence. 
Figure~\ref{AChevSTF-tra} is distinctive because PC2 is strongly associated with $p$ while PC1 remains more closely associated with $q$, demonstrating a clear separation of index effects across the first two components. 
Overall, the results identify $q$ as the consistently dominant index for PC1, whereas PC2 captures figure-dependent secondary variation linked to either $p$ or $q$. This component-specific behavior supports interpreting the reduced representation in terms of distinct, rather than uniformly coupled, index contributions.

Four of the 18 pairs of Pearson correlation coefficients in Table~\ref{tab:pca_correlations} are identified by stars. In a starred pair,
the specific principal component 
\begin{itemize}
\item decreases  as either $p$ increases/decreases or $q$ decreases/increases or
\item increases  as either $p$ increases/decreases or $q$ decreases/increases.
\end{itemize}
In an unstarred pair,
the specific principal component 
\begin{itemize}
\item decreases  as either $p$ increases/decreases or $q$ increases/decreases or
\item increases  as either $p$ increases/decreases or $q$ increases/decreases.
\end{itemize}
Whereas all nine pairs are unstarred for PC1, only five of the nine pairs are unstarred for PC2.

The PCA results show that $q$ is consistently associated with the dominant mode of variation (PC1) in the DIGP maps, whereas the secondary mode (PC2) exhibits figure-dependent association with either $p$ or $q$. The high cumulative variance explained by the first two or three principal
components indicates that the DIGP maps in each figure should be interpreted as a coordinated data family rather than as independent from each other.

\section{Conclusion}\label{sec:conc}
This paper has developed and illustrated the double-indexed geometric phase (DIGP) as a generalized phase-dependent observable for 
situations in which a time-harmonic electromagnetic field can be represented through a plane wave.
Building on the Pancharatnam phase premised on   symmetric Poincar\'e spinors,  a family of 
 Poincar\'e spinors distinguished by two indexes $p\in\mathbb{R}$ and $q\in\mathbb{R}$ is introduced.
 This family of spinors naturally engenders a corresponding family of double-indexed overlaps leading to the
 DIGPs.  The Pancharatnam phase is recovered as a special case, while the broader DIGP framework supplies complementary information about 
 polarization-state evolution that is not necessarily accessible through intensity-based observables.

Several symmetries   of the DIGP were established. Sign reversal of $p$  leaves the DIGP unchanged, whereas sign reversal
of $q$  introduces a correction proportional to the difference between the longitudes of the two plane waves being compared. The DIGP is periodic in $q$, with a period governed by that longitudinal separation. For closed loops, the DIGP retains a geometric interpretation: after mapping the original geographic coordinates into $p$-dependent effective geographic coordinates, the closed-loop DIGP equals one-half of the  solid angle subtended at the center of the
Poincar\'e sphere. These results show that the two indexes delineate systematic alternative projections, or spinor gauges, of the same polarization-state evolution
from that of the incident plane wave to the reflected/transmitted/scattered wave.

Two  examples of plane-wave scattering
demonstrate the potential practical value of DIGPs. First, the DIGP maps for reflection and transmission by chiral sculptured thin films  preserve the signatures of the circular Bragg phenomenon while revealing substantially finer structure than reflectance maps and transmittance maps do. Rapid DIGP variations occur near the circular Bragg regime and Fabry--P\'erot resonances, whereas smoother variations appear elsewhere. The maps generated with different  $(p,q)$ pairs form a multiparameter phase tomography  that may reveal the morphological and electromagnetic parameters of the thin film. A central  $90\deg$-twist defect produces additional phase mixing, sharper local contrasts, and a rearrangement of the spectral-angular features. These effects can remain pronounced even when the corresponding changes in conventional intensity data are modest, demonstrating the potential sensitivity of the DIGP to internal structural defects. A collection of DIGP maps should perhaps be treated as a coordinated data family rather than as independent observables. Similar conclusions followed
 asymmetric chevronic sculptured thin films, but with one surprising demonstration: whereas the reflectance and transmittance maps were affected prosaically
 by a central  $90\deg$-twist defect, the DIGP maps were affected dramatically.
 
 Second, the example of plane-wave scattering by a three-dimensional object
provides an especially clear contrast with conventional scattering measures. For a homogeneous isotropic chiral sphere, the differential scattering efficiency is strongly concentrated in the forward direction and is nearly featureless over much of the polar domain and all of the azimuthal domain. The direction-dependent DIGP, by contrast, exhibits pronounced dependence on both polar and azimuthal angles, including bands, lobes, extremums, and sign reversals. It therefore exposes polarization-state information hidden by the angular distribution of scattered intensity.

Taken together, these results indicate that the DIGP is best regarded as a complementary polarization-state observable rather than a replacement for reflectance, transmittance, or differential scattering efficiency. Its principal advantage is the retention of detailed phase-based information about the evolution of the polarization state, including information associated with   anisotropy and bianisotropy, structural chirality, resonances, and defects. The richness of DIGP maps suggests several prospective applications. For inverse-scattering
applications,  DIGP maps from polarimetric measurements could be compared with libraries generated by diverse analytical and computational techniques
to infer morphological and constitutive parameters. Thus, the DIGP concept could support nondestructive inspection of layered coatings, defect detection in chiral photonic structures, as well as characterization of metamaterials, metasurfaces, and other polarization-state-selective devices. It may also enhance optical sensing when a small structural change produces a weak intensity response but a significant geometric-phase response \cite{MLjosab2026}.

Further applications may arise in polarization-resolved radar and lidar, biomedical imaging, atmospheric-particle characterization, and remote sensing of structured or chiral media. A family of DIGP maps could serve as input to statistical inference, dimensional-reduction, or machine-learning methods for automated classification of materials and morphologies. Realizing these possibilities will require experimental phase-sensitive polarimetry, robust calibration, careful treatment of phase singularities and branch ambiguities, and systematic assessment of noise and multiple-scattering effects. Subject to these developments, the DIGP concept provides a promising extension of geometric-phase methods and a potentially powerful route toward polarization-state-based electromagnetic characterization with enhanced structural sensitivity.

\section*{Appendix 1: Vector Spherical Wavefunctions}
The vector spherical wavefunctions  \cite{Stratton1941,MF1953}
\begin{equation}
{\#M}\eomn\three(\ko{\#r}) = 
h_n\one(\ko r)\les
\utheta \,f\eomn(\theta,\phi) 
-\uphi \,g\eomn (\theta,\phi)\ris\,
\label{M3-def}
\end{equation}
and
\begin{eqnarray}
\nonumber
&&
{\#N}\eomn\three(\ko{\#r}) =
 \ur \,n(n+1) P_n^m(\cos\theta)\frac{h_n\one(\ko r)}{\ko r}
{\lec\begin{array}{c}{\cos(m\phi)}\\{\sin(m\phi)}\end{array}\ric}
\\[5pt]
&&\qquad+ \frac{\psi_n\three(\ko r)}{\ko r}
\les\utheta \,g\eomn(\theta,\phi)+\uphi\,f\eomn(\theta,\phi)\ris
\label{N3-def}
\end{eqnarray}
are regular at infinity. In these expressions,
\begin{subequations}
\begin{equation}
f\eomn(\theta,\phi)=\mp\pinm(\theta)
{\lec\begin{array}{c}{\sin(m\phi)}\\{\cos(m\phi)}\end{array}\ric}\,,
\label{def-fsmn}
\end{equation}
\begin{equation}
g\eomn(\theta,\phi)=\taunm(\theta)
{\lec\begin{array}{c}{\cos(m\phi)}\\{\sin(m\phi)}\end{array}\ric}\,,
\label{def-gsmn}
\end{equation}
\begin{equation}
\pinm(\theta)=\frac{m P_n^m(\cos\theta)}{\sin\theta}\,,
\end{equation}
\begin{equation}
\taunm(\theta)=\frac{dP_n^m(\cos\theta)}{d\theta}\,,
\end{equation}
\begin{equation}
 \psi_n\three(w)= \frac{d}{dw}\left[w\,h_n\one(w)\right]\,,
\end{equation}
\end{subequations}
 $h_n\one(\.)$ as the spherical Hankel function of the first kind and order $n$,
and $P_n^m(\.)$ is the associated
Legendre function of order $n$ and degree $m$.

 \section*{Acknowledgments}
AL  was partially supported
by the Evan Pugh University Professorships Endowment at Penn State.
TGM and AL were partially supported by EPSRC
(grant number APP68512).


\begin{thebibliography}{999}
\bibitem{footnote1}{The Pancharatnam phase is also called
the Pancharatnam--Berry phase in optics literature.}
\bibitem{footnote3}
{If the closed-loop overlap were defined as $\upsilon_{1231}=\upsilon_{12}+\upsilon_{23}+\upsilon_{31}$, it would not be invariant
under arbitrary rephasing of the spinors, i.e., if $\les\psis_\ell\ris\to \exp(i\chi_\ell)\les\psis_\ell\ris$, $\ell\in\lec1,2,3\ric$.}
\end{thebibliography}
\end{document}